\documentclass[preprint]{aastex}

\usepackage{color,amsmath,url}
\newcommand\kmps{\mbox{$\rm km\,s^{-1}$}}
\newcommand\vLSR{\mbox{$v_{\rm LSR}$}}
\newcommand\vsys{\mbox{$v_{\rm sys}$}}
\newcommand\Msun{\mbox{$M_\sun$}}
\newcommand\Lsun{\mbox{$L_\sun$}}
\newcommand\Mspy{\mbox{$\Msun\rm\,yr^{-1}$}}
\newcommand\mJpb{\mbox{mJy\,beam$^{-1}$}}

\shorttitle{CIT 6}
\shortauthors{KIM et al.}

\begin{document}
\title{High-resolution CO observation of the carbon star CIT 6 
revealing the spiral structure and a nascent bipolar outflow}
\author{Hyosun Kim\altaffilmark{1,2,3}}
\author{Sheng-Yuan Liu\altaffilmark{1}}
\author{Naomi Hirano\altaffilmark{1}}
\author{Ronny Zhao-Geisler\altaffilmark{4,1}}
\author{Alfonso Trejo\altaffilmark{1}}
\author{Hsi-Wei Yen\altaffilmark{1}}
\author{Ronald E. Taam\altaffilmark{1}}
\author{Francisca Kemper\altaffilmark{1}}
\author{Jongsoo Kim\altaffilmark{2}}
\author{Do-Young Byun\altaffilmark{2}}
\author{Tie Liu\altaffilmark{2}}
\altaffiltext{1}{Academia Sinica Institute of Astronomy and Astrophysics, 
  P.O. Box 23-141, Taipei 10617, Taiwan; hkim@asiaa.sinica.edu.tw}
\altaffiltext{2}{Korea Astronomy and Space Science Institute, 776, 
Daedeokdae-ro, Yuseong-gu, Daejeon 305-348, Korea}
\altaffiltext{3}{EACOA fellow}
\altaffiltext{4}{National Taiwan Normal University, Department of Earth 
Sciences, 88 Sec.~4, Ting-Chou Rd, Wenshan District, Taipei 11677, Taiwan}

\begin{abstract}
CIT 6 is a carbon star in the transitional phase from the asymptotic 
giant branch (AGB) to the protoplanetary nebulae (pPN). Observational 
evidences of two point sources in the optical, circumstellar arc 
segments in an HC$_3$N line emission, and a bipolar nebula 
in near-infrared provide strong support for the presence of a binary 
companion. Hence, CIT 6 is very attractive for studying the 
role of companions in the AGB-pPN transition. 
We have carried out high resolution $^{12}$CO $J$=2--1 and $^{13}$CO 
$J$=2--1 observations of CIT 6 with the Submillimeter Array 
combined with the Submillimeter Telescope (single-dish) data. 
The $^{12}$CO channel maps reveal a spiral-shell pattern connecting 
the HC$_3$N segments in a continuous form, and an asymmetric outflow 
corresponding to the near-infrared bipolar nebula. Rotation of the 
$^{12}$CO channel peak position may be related to the inner 
spiral winding and/or the bipolar outflow. 
An eccentric orbit binary is suggested for the presences 
of an anisotropic mass loss to the west and a double spiral pattern. 
The lack of interarm emission to the west may indicate a feature 
corresponding to the periastron passage of a highly eccentric orbit of 
the binary. Spatially-averaged radial and spectral profiles of 
$^{12}$CO $J$=2--1 and $^{13}$CO $J$=2--1 are compared with simple 
spherical radiative transfer models, suggesting a change of 
$^{12}$CO/$^{13}$CO abundance ratio from $\sim30$ to $\sim50$ inward 
in the CSE of CIT 6. 
The millimeter continuum emission is decomposed into extended dust 
thermal emission (spectral index $\sim-2.4$) and compact 
emission from radio photosphere (spectral index $\sim-2.0$). 
\end{abstract}


\section{INTRODUCTION}\label{sec:int}

Beyond the main-sequence of a stellar population, a star with an initial 
mass less than 8\,\Msun\ evolves to the red giant phase at which it 
ejects most of its mass through a slow and dense wind ($3-30\,\kmps$). 
The intense mass loss process reaches its climax at the tip of the 
asymptotic giant branch (AGB) with a rate of $10^{-7}-10^{-4}\,\Mspy$ 
forming an extensive circumstellar envelope (CSE, $10^{16}$--$10^{18}
\rm\,cm$). Such an evolutionary phase leads to the obscuration of the 
innermost envelope region near the stellar surface. Consequently, 
information on the onset of dust formation and growth, the subsequent 
wind acceleration, the central engine governing the (a)symmetric shapes 
of the entire CSE, and the connection of these features with binary 
interactions and/or magnetic fields is very limited. 

The morphological diversity of the CSEs in evolved stellar phases is 
commonly assumed to be closely related to the possible binary nature 
of the system. In contrast to quasi-spherical shapes of CSEs at the 
AGB, the majority of planetary nebulae (PNe) or protoplanetary nebulae 
(pPNe) bear bipolar or multipolar structures, often observed in ionized 
gas or dust scattered light \citep[e.g.,][]{cor95,bal02,sah07}. 
Such bipolarity is a common feature of sources possessing accretion 
disks such as the protoplanetary jets coexisting with the accretion 
disks surrounding young stellar objects. In the outflowing CSE of 
evolved giant stars, another gravitational potential (e.g., companion) 
is hypothesized to assemble material in the form of circumstellar or 
circumbinary disk to feed the observed bipolar outflow. The accretion 
disks may form during the AGB phase in {\em close} binary systems 
\citep[e.g.,][]{the93,hua13}.

{\em Wide} binary systems also induce, albeit less vigorous, asymmetry 
in the form of spiral patterns. It is now theoretically well understood 
that binary orbital motion leads to a spiral-shell pattern propagating 
over the entire CSE \citep[e.g.,][]{sok94,mas99,moh12,kim12a,kim12b}. 
The consequent wind anisotropy, caused by the binary orbital motion 
and the gravitational focusing of the wind material onto the companion, 
leads an overall flattened geometry of the CSE \citep{hug09,kim12b}. 
The binary-induced pattern and geometry of CSE preserve the records of 
binary influence on the wind outflow. 
A well-defined spiral pattern of AFGL 3068 found from dust scattered light 
and the binary separation inferred from near-infrared adaptive optics images 
\citep{mor06,mau06} were utilized to derive its binary stellar and orbital 
properties \citep{kim12c}. The non-continuous shells within the CSE of CIT 
6, which appeared in a molecular line emission \citep{cla11}, were suggested 
to be a spiral-shell pattern when viewed with a large inclination from the 
orbital plane \citep{kim13}. Such an interpretation may also be applicable 
to the multiple rings and arcs often found in the CSEs of AGB stars and pPNe, 
e.g., IRC+10216 \citep{mau99,dec15,cer15} and Mira \citep{may11,ram14}. 
In addition, within this framework, the wind properties derived from the 
circumstellar spiral pattern facilitated constraints on the mass loss 
history of R Sculptoris \citep{mae12}. Finally, the elliptical shapes 
of CSEs of AGB stars in an optical imaging survey were suggested as a 
probe to infer that a significant fraction of AGB stars may have hidden 
binary companions \citep{mau13}.

In fact, substantial pPNe may reveal evidence of the shape transition 
from outer rings and arcs formed in the CSE during the previous AGB 
phase to the inner region showing bipolar lobes developed by new vigorous 
ejections \citep[e.g., Egg nebula,][]{sah98}. If the ring-like patterns 
originated from binary orbital motion, the typical dynamical timescales 
of the observed patterns (100\,--\,1000 years) correspond to the orbital 
periods of wide binary systems with the companions locating at a few tens 
to hundred AU. Such separations between binary stars are far larger than 
radius of wind acceleration zone so that the wind speed is sufficiently 
fast in the vicinity of the companion (if it is in a circular orbit), 
precluding formation of an accretion disk and consequently a bipolar 
outflow. Whether a {\em wide} binary can create a bipolar outflow has 
not been well studied. The onset of the asymmetric outflow, the role of 
a binary companion on it, and ultimately the mass loss process may be 
addressed by sub-arcsecond resolution observations toward the sources 
in this phase transition from the AGB to pPN. 

The extreme carbon star CIT 6 is an excellent target to study the role of 
a companion on the transition from a quasi-spherical shell-like structure 
(over $\sim20\arcsec$) found in a molecular line emission map \citep{cla11} 
to a nascent bipolar structure ($\la2\arcsec$) observed in the near-infrared 
polarization image \citep{sch02}. The companion candidate revealed in optical 
images is $\la0\farcs2$ away from the carbon star at a position angle (PA, 
from north to east) of $\sim10\arcdeg$ \citep{mon00}. 
At a distance of 400\,pc \citep[and references therein]{coh96} derived 
by using a period-luminosity relation from the pulsation period of the 
carbon star ($\sim1.65$\,yr), the projected binary separation corresponds 
to $\la80$\,AU. The observed flux at 4500\,\AA\ suggests the candidate 
companion to be a main-sequence star of mass $\sim1$--\,2\Msun\ \citep{sch02}. 
On the other hand, the effective temperature $\sim2800$\,K \citep{coh79} and 
bolometric luminosity $\sim10^4\,\Lsun$ \citep{lou93} locate the carbon star 
CIT 6 at the tip of AGB of a star with an initial mass of $\sim2$\,\Msun\ 
(current mass $\la1\Msun$) based on the stellar evolutionary model with solar 
metallicity using the code provided by \citet{hur00}. These 
photometric masses agree with the masses derived from a separate, parameter 
space analysis based on binary dynamics and the encoded circumstellar pattern 
\citep{kim13}. 

The circumstellar pattern of CIT 6 was observed in the HC$_3$N 
$J$\,=\,4--3 line emission at a high resolution of $\sim0\farcs7$ 
with the Karl G.\ Jansky Very Large Array (JVLA). The pattern was 
modeled with spherical shells, resulting in a necessary displacement of 
shell centroids along the radius and across the line velocity channels 
(\citealp{cla11}; see also \citealp{cha12} for various HC$_3$N and 
HC$_5$N lines of CIT 6). This departure from spherical symmetry is 
consistent with the spiral-shell model induced by a binary motion, and 
\citet{kim13} indeed presented a spiral-shell model reproducing the 
observed HC$_3$N pattern in all channels. Previously, \citet{din09} 
speculated a spiral structure of CIT 6 from another HC$_3$N $J$\,=\,5--4 
emission map at a lower resolution of $\sim1\farcs5$, which is now resolved 
into a few windings of the modeled spiral-shell pattern. 
Nevertheless, the pattern geometry was left incomplete, because the 
observed HC$_3$N pattern is broken with an absence of emission 
to the west and in the central $\sim3\arcsec$ region. Verifying the 
pattern shape with a chemically stable molecule such as $^{12}$CO is 
necessary to infer the companion of CIT 6, to examine the binary 
characteristics, and to obtain a clue of the kinematic connection of 
the spiral pattern to the observed near-infrared bipolar nebula in 
the innermost region where the HC$_3$N emission was absent.

In this paper, we report the Submillimeter Array\footnote{The Submillimeter 
Array is a joint project between the Smithsonian Astrophysical Observatory 
and the Academia Sinica Institute of Astronomy and Astrophysics, and is 
funded by the Smithsonian Institution and the Academia Sinica.} (SMA) 
observations toward CIT 6 in the $^{12}$CO $J$\,=\,2--1 and $^{13}$CO 
$J$\,=\,2--1 lines and the 226.2\,GHz and 350.6\,GHz continuum emission 
at sub-arcsecond resolutions. Submillimeter Telescope (SMT) grid-mapping 
observations for the same lines are combined with the SMA data to recover 
the extended CSE emission. We describe the observations and calibrations 
in Section\,\ref{sec:obs}. 
The results are presented in Sections\,\ref{sec:ove}
and \ref{sec:fin}
for the overall and fine structures in the CSE of CIT 6, respectively. 
Further discussion follows in Section\,\ref{sec:dis}. 
We summarize our findings in Section\,\ref{sec:sum}, and conclude in 
Section\,\ref{sec:ccs}.

\section{OBSERVATIONS}\label{sec:obs}
\subsection{SMA Observations}\label{sec:sma}

The $^{12}$CO $J$\,=\,2--1 and $^{13}$CO $J$\,=\,2--1 lines and the adjacent 
continuum (at 226.2 GHz) have been simultaneously observed with the SMA in 
four different configurations (see Table\,\ref{tab:sma}). These configurations 
provided the projected baselines ranging from $\sim5$\,m to $\sim500$\,m. 
The observations were carried out on five different dates. On 2013 Jan.\ 08 
and Feb.\ 06 for extended and very extended configurations, two receivers 
were simultaneously setup with one receiver at $\sim226$\,GHz covering 
$^{12}$CO $J$\,=\,2--1 and $^{13}$CO $J$\,=\,2--1, and the other receiver 
at $\sim350$\,GHz. Each receiver has 2\,GHz IF-bandwidth. On 2013 Dec.\ 
12, 2014 Feb.\ 09, and 2014 Mar.\ 11, however, one 230\,GHz receiver was 
setup with 4\,GHz IF-bandwidth. 
Several lines were detected, but were spatially unresolved; hence, in this 
paper, we focus on the $^{12}$CO $J$\,=\,2--1 and $^{13}$CO $J$\,=\,2--1 
lines. The primary beam sizes of the SMA antenna are $\sim55\arcsec$ at 
230\,GHz and $\sim36\arcsec$ at 350\,GHz, respectively. 

The visibility data were calibrated with the standard procedure using the MIR 
package\footnote{See \url{https://www.cfa.harvard.edu/~cqi/mircook.html}.} 
with two quasars 0854+201 and 0927+390 as the gain (phase and amplitude) 
calibrators, and a quasar, either 3C84 or 3C279, as the bandpass calibrator. 
The flux calibrations were mainly based on the SMA flux monitoring program of 
a large number of quasars, tabulated in the Tools section of the SMA Observer 
Center \citep{gur07}; when Titan or Callisto is available, the fluxes 
of the observed quasars measured based on the planetary object were 
consistent with the monitoring records. 

The MIRIAD packages were used for the inverse Fourier transformation 
and image cleaning from the calibrated visibility data to the final 
channel maps. 
The synthesized beam size of the $^{12}$CO $J$\,=\,2--1 line emission 
map is $0\farcs56\times0\farcs44$ with a PA of $40\arcdeg$ with the 
uniform weighting of the visibility data from all configurations 
listed in Table\,\ref{tab:sma}. The velocity interval was set to be 
1\,\kmps\ in the imaging step. The noise level $\sigma$ estimated 
from the line-free channels is 24\,\mJpb\ (2.2\,K) at the velocity 
resolution of 1\,\kmps.
The continuum map at 226.2\,GHz was attained by averaging over the 
line-free channels of both sidebands (separated by 10\,GHz) with 
uniform weighting of the data from all SMA configurations. 
Its synthesized beam has a size of $0\farcs54\times0\farcs43$ with 
a PA of 40\arcdeg, and the noise level is 0.75\,\mJpb\ (76\,mK). 
The 350.6\,GHz continuum imaging is characterized by a synthesized 
beam of $0\farcs36\times0\farcs30$ with a PA of 25\arcdeg\ and a 
noise of 3.6\,\mJpb\ (330\,mK).

\subsection{SMT Observations}\label{sec:smt}

The SMT grid-mapping observations have been carried out to fill in the zero 
and short baselines of the $^{12}$CO $J$\,=\,2--1 and $^{13}$CO $J$\,=\,2--1 
lines using the 1.3\,mm sideband-separating receivers with dual-polarization. 
The integration per pointing of the grid-mapping is 3\,min for $^{12}$CO and 
6\,min for $^{13}$CO (see Table\,\ref{tab:smt}). 
Deep single-point observations toward the central position were additionally 
performed; the total integration for $^{12}$CO central pointing is 11\,min, 
and 392\,min for $^{13}$CO. 
A 1024-channel (2$\times$512) backend with a 1\,MHz filterbank was employed. 
The On/Off-integrations were repeated to correct the sky background variation 
via position-switching for the grid-mapping mode and beam-switching for the 
single-point mode. Pointing corrections were monitored by observations of 
Jupiter. The beam size of a single pointing is $\sim33\arcsec$ at 1.3\,mm.

Using the CLASS software package in GILDAS\footnote{GILDAS is developed 
and distributed by the Observatoire de Grenoble and IRAM.}, we subtracted 
the linear baseline fits, and co-added the individual scans to achieve 
the final spectra. The temperature scales, $T_{\rm A}^*$, of $^{12}$CO 
and $^{13}$CO lines are 2.95\,K and 0.20\,K, respectively, at the peaks, 
comparable (with $<$\,20\,\% uncertainty) to the values in earlier SMT 
single-pointing observations \citep[2.49\,K for $^{12}$CO and 0.22\,K 
for $^{13}$CO, by][]{mil09}. The main-beam temperatures were derived 
by $T_{\rm B}=T_{\rm A}^*/\eta_{\rm mb}$ with the main-beam efficiency 
$\eta_{\rm mb}$ ($=0.74$). The $1\sigma$ noise levels of the grid-mapping 
data (in $T_{\rm B}$) are 17\,mK and 15\,mK for the $^{12}$CO and $^{13}$CO 
maps, respectively. The noise levels of central pointings with the long 
integrations are 11\,mK for the $^{12}$CO and 3\,mK for the $^{13}$CO.

\subsection{Joint Deconvolution of SMA and SMT Data}\label{sec:all}

\begin{figure*} 
  \plotone{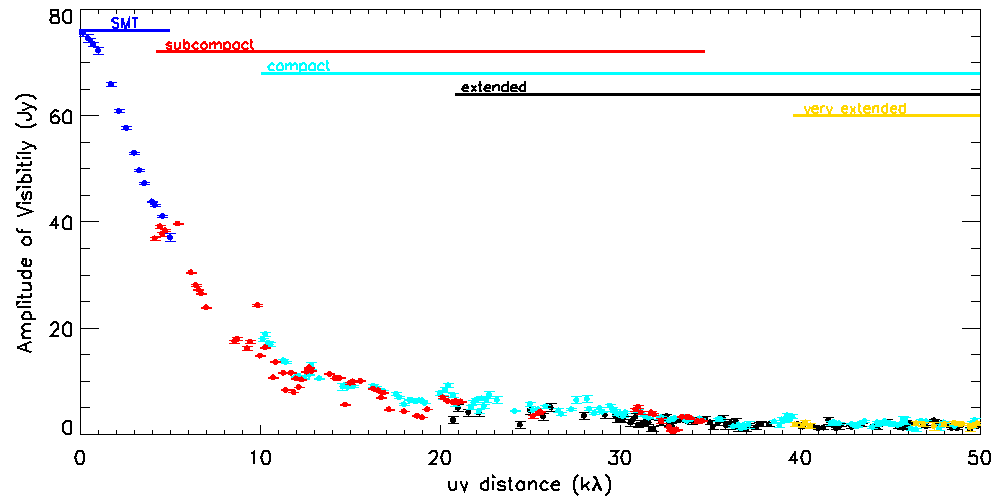}
  \caption{\label{fig:uvl}
    Visibility amplitude of $^{12}$CO $J$=2--1 line emission from CIT 6 
    (integrated over 40\,\kmps) along the $uv$ distance of the SMT (blue; 
    2015 Jan.\ 07 and 17) and SMA's subcompact (red; 2014 Feb.\ 09 and 
    Mar.\ 11), compact (cyan; 2013 Dec.\ 12), extended (black; 2013 
    Jan.\ 08) and very extended (yellow; 2013 Feb.\ 06) configurations. 
    The visibility data are shown with 350 bins in the $uv$ distance up to 
    50\,k$\lambda$. The full $uv$ distance coverage is up to 400\,k$\lambda$. 
    }
\end{figure*}

We used MIRIAD to combine the SMA interferometric and SMT single-dish data 
following the procedure described by \citet{kod11}. Visibilities from the 
SMT single-dish images are generated by (1) deconvolving the SMT maps with 
their Gaussian beams (FWHMs of $\sim33\arcsec$ and $\sim34\arcsec$ for 
$^{12}$CO and $^{13}$CO frequencies, respectively), (2) attenuating these 
deconvolved SMT images with the SMA primary beams (FWHMs of $\sim55\arcsec$ 
and $\sim57\arcsec$, respectively), and (3) generating visibilities from the 
above (SMA primary beam attenuated) maps. 
We set the integration time per visibility equivalent to be the one per SMA 
visibility. The number of SMT visibilities was $3\times10^4$; varying this 
number within a factor of a few does not change the final image as far as 
the uniform weighting is employed. These SMT visibilities are combined with 
the SMA visibilities from all configurations and Fourier-transformed to 
dirty images by the MIRIAD's {\tt invert} task. Figure\,\ref{fig:uvl} 
shows the flux consistency of the visibility data from the SMT and 
the different configurations of SMA. 

After the deconvolution process with the MIRIAD's {\tt clean} task, we 
restored the clean components with the Gaussian beams that have the same 
solid angles as their corresponding dirty beams. \citet{kod11} emphasized 
the importance of this process for conserving the total flux, in particular, 
when sources with extended low level emission are involved. 
Conventionally in a restored image, the clean components are convolved with 
a synthesized beam (defined as a Gaussian function fitting the main lobe of 
the dirty beam), and the residuals (in units of the dirty beam solid angle) 
are added back. Departure of the dirty beam from the Gaussian shape leads to 
inconsistency in the beam solid angle between the dirty beam and the derived 
synthesized beam, resulting in an artificial mismatch of the total flux. 
The conventional method works well in the cases of compact sources, where 
the cleaning process results in negligible residuals. In our case, because 
of the extended nature of the CSE and the non-Gaussian shape of the dirty 
beam, we defined new Gaussian beams having the same solid angles as the 
dirty beams, ensuring the total fluxes to be conserved after cleaning 
and restoring the images. The PAs and beam axis ratios are set to the ones 
for the main lobes of dirty beams. 

Finally, the combined images are corrected for the SMA primary beam 
attenuation. More detailed description for combining the single-dish and 
interferometer data can be found in \citet{kod11}. The final images of 
$^{12}$CO and $^{13}$CO have the beam sizes of $2\farcs14\times1\farcs74$ 
and $2\farcs23\times1\farcs79$, respectively, with the PA of 37\arcdeg. 
The noise levels are 25\,\mJpb\ (150\,mK) for the $^{12}$CO map and 
22\,\mJpb\ (140\,mK) for the $^{13}$CO map.

\section{FEATURES OF OVERALL STRUCTURE}\label{sec:ove}
\subsection{Continuum Emission: Position and Flux of the Radio Photosphere}
\label{sec:con}

\begin{figure*} 
  \plotone{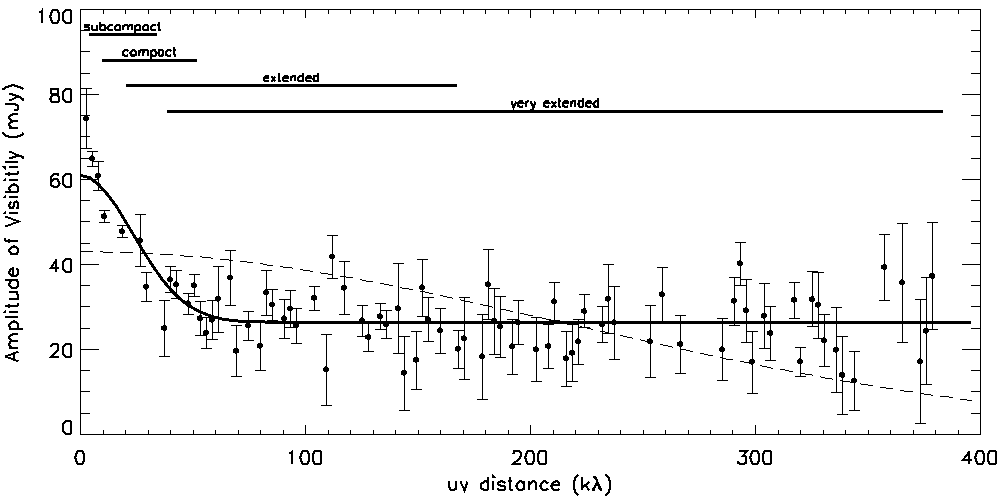}
  \caption{\label{fig:uvc}
    Visibility amplitude at 226.2\,GHz continuum along the $uv$ distance 
    of the SMA, averaged per 2.65\,k$\lambda$-width bin. Solid curve 
    represents our two-component (one point and one Gaussian) model 
    (see Table\,\ref{tab:con}). For comparison, a single Gaussian model 
    is overlaid by a dashed curve. 
  }
\end{figure*}

The position and flux of the continuum are determined by the {\tt uvfit} 
task of MIRIAD. 
In both 226.2\,GHz and 350.6\,GHz continuum, the visibility amplitudes 
are fairly flat in the $uv$ distance (i.e., projected baseline length in 
wavelengths) range of 40\,--\,400\,k$\lambda$ (see Figure\,\ref{fig:uvc}), 
implying a spatially unresolved source. In this $uv$ distance range, we have 
tried to find the best point, Gaussian, or disk source model, resulting in 
the flux and positional offset almost independent on the models. Moreover, 
the resulting Gaussian and disk sizes are comparable to the sizes in our 
trial fit for a phase calibrator, essentially a point source; 
therefore we adopt the point source model for the visibility data at 
40\,--\,400\,k$\lambda$. 

\begin{figure*} 
  \plotone{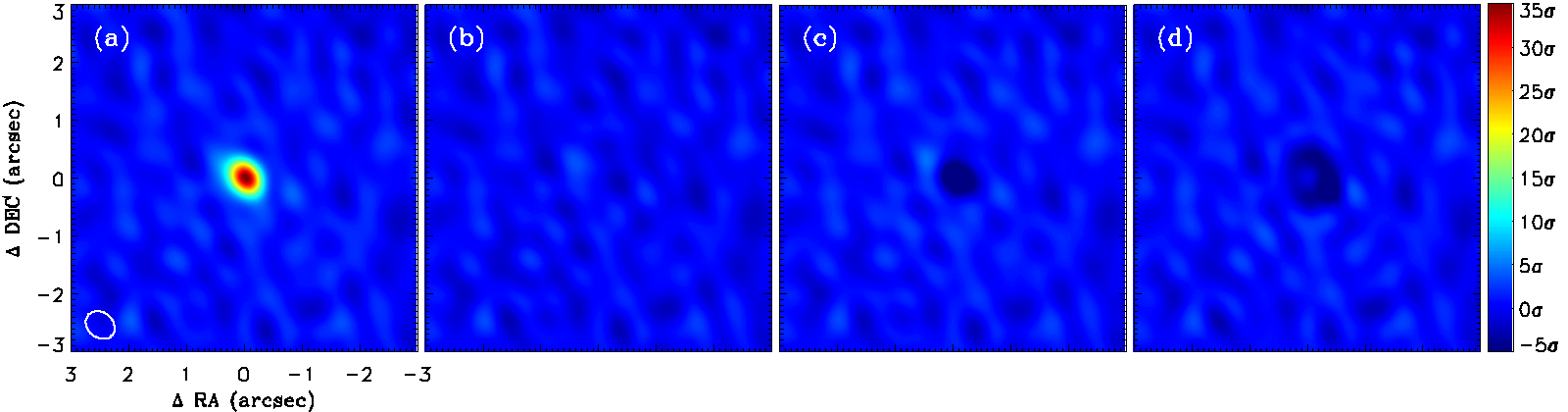}
  \caption{\label{fig:con}
    (a) Map of the continuum emission at 226.2\,GHz taken with the SMA 
    all configurations. Residuals after subtracting (b) two-component 
    (point source + Gaussian) model, (c) one point source model, and 
    (d) one Gaussian model. The synthesized beam size is indicated by 
    a white contour in (a). The colorbar is labeled in a linear scale 
    in units of the noise level $\sigma=0.75\,\mJpb$.
  }
\end{figure*}

The 226.2\,GHz continuum visibility amplitudes from the subcompact and 
compact configurations of SMA (4\,--\,50\,k$\lambda$) are higher and 
increase toward the shorter baselines. The visibility data from all SMA 
configurations are best fit with a two-component model in a combination 
of one point source and one extended component as denoted by a solid 
curve in Figure\,\ref{fig:uvc} (see Table\,\ref{tab:con}). 
The residual image of the two-component model indeed shows a random 
noise distribution (Figure\,\ref{fig:con}(b)), in contrast to the 
one-component model where unexpected structures remain in the residuals 
(Figure\,\ref{fig:con}(c)--(d)). Given the uncertainty of visibility 
data, it is currently difficult to differentiate the Gaussian and disk 
models for the extended component. 

\begin{figure} 
  \plotone{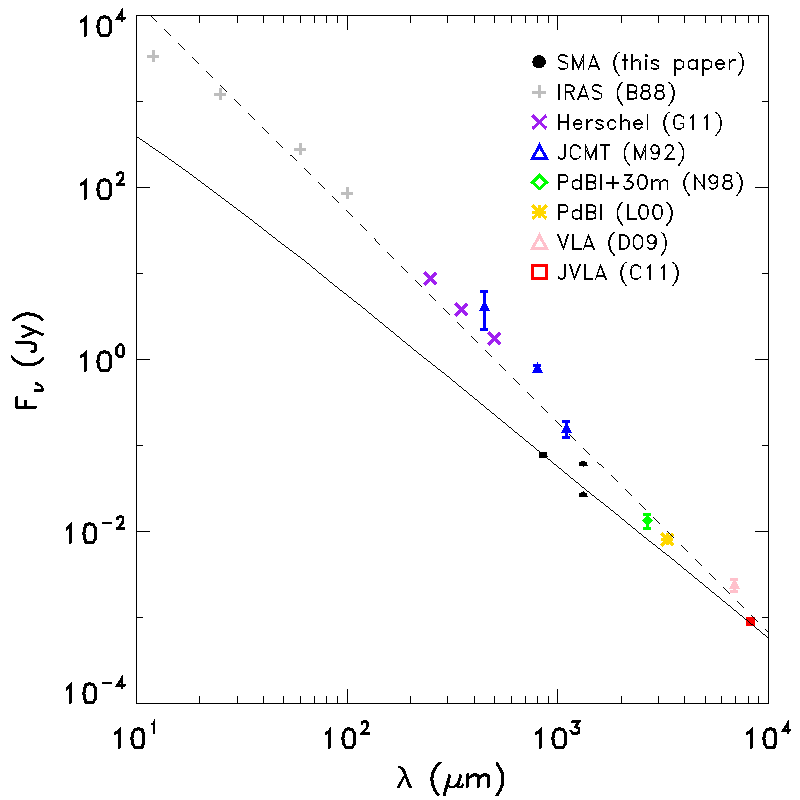}
  \caption{\label{fig:sed}
    Spectral energy distribution of CIT 6 from infrared to centimeter 
    wavelengths ($\sim10\,\mu$m to $\sim1$\,cm). 
    The SMA flux at 226.2\,GHz (1.3\,mm) and 350.6\,GHz (0.8\,mm) are 
    measured via visibility fits. Solid line represents thermal black 
    body radiation curve from the suggested radio photosphere of CIT 6 
    with a temperature of 2000\,K and a diameter of 8\,AU (20\,mas). 
    Dashed line shows a linear fit to the accumulated data from the 
    literature, resulting in a spectral index of $-2.4$. 
    The references B88, G11, M92, N98, L00, D09, and C11 represent 
    \citet{bei88}, \citet{gro11}, \citet {mar92}, \citet{ner98}, 
    \citet{lin00}, \citet{din09}, and \citet {cla11}, respectively. 
  }
\end{figure}

Figure\,\ref{fig:sed} presents the comparison of the SMA continuum flux at 
226.2\,GHz and 350.6\,GHz with various measurements known in the literature 
at different frequencies. 
The 226.2\,GHz and 350.6\,GHz flux of the unresolved continuum component 
(together with a 36.5\,GHz continuum flux from another subarcsec-resolution 
observation with the JVLA) is described by a spectral index $n=-2$, where 
$n$ is defined by $F_\nu\propto\lambda^n$. This spectral index is expected 
for thermal blackbody radiation in the Rayleigh-Jeans limit. The compact 
nature and the spectral index close to $-2$ in the millimeter and centimeter 
wavelengths may indicate that the unresolved component of the SMA continuum 
emerges from a ``radio photosphere'' surrounding the stellar (optical) 
photosphere, which is introduced by \citet{rei97} for long-period variables. 
\citeauthor{rei97} showed that the radio photosphere is about twice the size 
of the optical photosphere. 

With an effective temperature ($\sim2800$\,K) of CIT 6 from the observed 
Na I D line at 5893\AA\ \citep{coh79} and its bolometric luminosity $\sim
10^4\,\Lsun$ from the integrated infrared emission \citep{lou93}, the 
Stefan-Boltzmann law $L_*/\Lsun=(R_*/R_\sun)^2(T_*/T_\sun)^4$ implies the 
stellar (optical) photosphere radius $R_*\sim2$\,AU. Adopting the radio 
photosphere to be twice the size of the stellar photosphere \citep{rei97}, 
we applied a diameter of 8\,AU (20 mas) and the corresponding radio 
photosphere brightness temperature of 2000\,K for a black body curve drawn 
by a solid line in Figure\,\ref{fig:sed}. This line is consistent with the 
SMA measurements for the unresolved continuum source. The radio photosphere 
size of CIT 6 is comparable to that of IRC+10216 (11\,AU) observed as a 
resolved disk at 43.3\,GHz continuum \citep{men12}.

The data from infrared to centimeter wavelengths are fitted to a spectral 
index of $-2.4$, confirming the result of \citet{mar92} via a fit over a 
wide wavelength range (0.5\,\micron\,--\,1\,mm). It presents the excess 
thermal emission from circumstellar dusts. 
The upper point at 226.2\,GHz (1.3\,mm) in Figure\,\ref{fig:sed} is the 
total flux of the two-component (point source and Gaussian) model given 
in Table\,\ref{tab:con}. If the dashed line fit provides a rough estimate 
for the total flux at 226.2\,GHz, albeit not obvious with the large 
deviation of the accumulated data, the extended continuum component of 
our SMA observations (34.5\,mJy, FWHM\,$\,\la4\arcsec$) recovers about 
50\% of the total thermal dust emission. There might exist more extended 
dust emission, missed in the SMA interferometric observations. 

\begin{figure} 
  \plotone{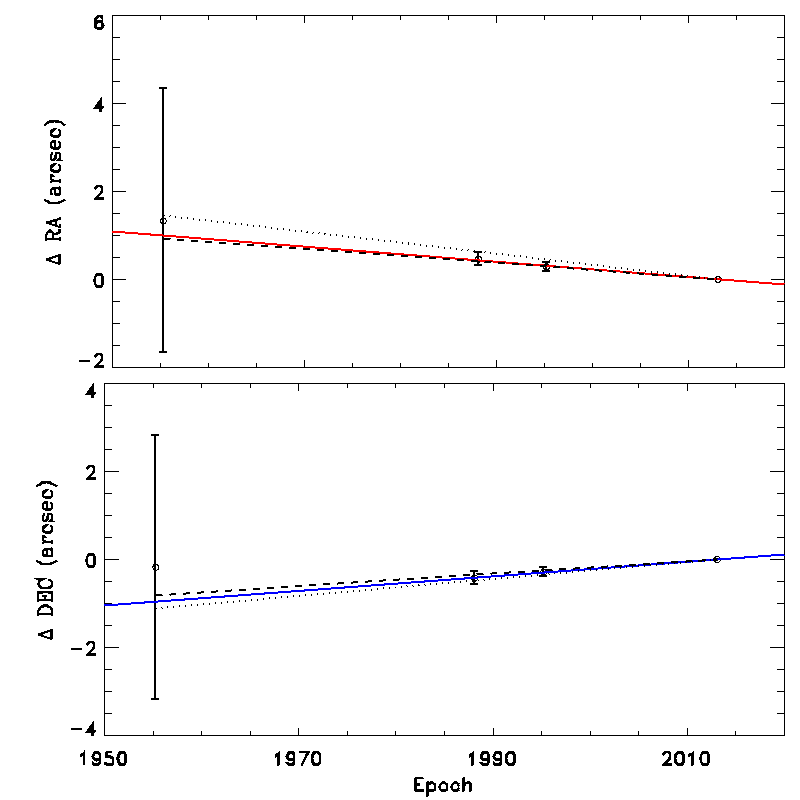}
  \caption{\label{fig:cen}
    Proper motion of CIT 6 based on the continuum emission at 226.2\,GHz
    from the SMA (epoch 2013.1), compared with the continuum emission at 
    90.7\,GHz \citep[in epoch 1995.2,][]{lin00}, the HCN maser \citep[in 
    epoch 1988.11]{car90}, and the optical image \citep[in epoch 1955.29]
    {cla87}. The colored solid lines present our proper motion measurement 
    by linear fitting of the four data points, and the dashed and dotted 
    lines indicate earlier measurements by \citet{mon00} and \citet{roe08}, 
    respectively.
  }
\end{figure}

The coordinate center in this paper corresponds to right ascension 
(RA) at 10$^{\rm h}$16$^{\rm m}$02\fs259 and declination (Dec) at 
$+$30\arcdeg34\arcmin19\farcs18 in J2000 coordinates, derived from 
the visibility fitting of the unresolved continuum emission of CIT 
6 at 226.2\,GHz. 
Figure\,\ref{fig:cen} shows the comparison of this position at the epoch 
2013.1 with the optical, HCN maser, and 90.7\,GHz continuum positions of 
CIT 6 at the epochs 1955.3, 1988.1, and 1995.2 \citep{cla87,car90,lin00}. 
From the center positions in the four measurements, we calculate the linear 
proper motion of ($-17\pm4$, $16\pm4$)\,mas\,yr$^{-1}$, which is in accord 
with those from the direct measures of positional shifts with respect to 
reference stars in optical and infrared images; ($-16\pm10$, $14\pm10$)\,%
mas\,yr$^{-1}$ \citep{mon00}, and ($-25\pm10$, $19\pm10$)\,mas\,yr$^{-1}$ 
\citep{roe08}. Therefore, CIT 6 is moving to the northwest direction with 
a projected speed of $\sim44\,\kmps$, which is also consistent with the 
speed derived from an astrospheric feature at $\sim15\arcmin$--$18\arcmin$ 
in far-ultraviolet images \citep{men12,sah14}.

\subsection{Spectral Profile}\label{sec:spe}

Figure\,\ref{fig:spe} illustrates the spectral profiles of $^{12}$CO 
$J$\,=\,2--1 and $^{13}$CO $J$\,=\,2--1 lines varying with integration 
radius. The $^{12}$CO spectra have a red-skewed doubly-peaked profile 
when integrated over the central region ($r\leq1\arcsec$) as shown in 
panel (a), a flat-topped profile over $\leq5\arcsec$ region in (b), and 
approximately parabolic profiles over larger $\sim33\arcsec$ and $\sim2
\farcm5$ regions in (c) and (d). These parabolic and flat-topped shapes 
of $^{12}$CO line profiles are typical for an optically thick source 
when it is spatially unresolved and resolved, respectively \citep{mor75}. 

In contrast, the $^{13}$CO spectral profiles have double peak features 
regardless of the integration areas, although the spectrum in (d) 
integrated over the entire SMT mapping area may be also interpreted as 
a flat-topped profile with the given noise. According to \citet{mor75}, 
an optically thin source produces a double peak shape in the spectral 
profile when it is spatially resolved, and a flat-topped profile when 
unresolved; hence, the $^{13}$CO line emitting region is likely similar 
to, or smaller than, the SMT mapping area (i.e., the integrated area for 
(d) panel). The $^{13}$CO envelope is generally smaller than the $^{12}$CO 
envelope because of the self-shielding capability of the abundant $^{12}$CO 
molecules against photodissociation by interstellar ultraviolet radiation 
field. Therefore, considering the nearly flat-topped shape of $^{13}$CO 
line together with the parabolic shape of $^{12}$CO line, our SMT mapping 
area ($\sim2\farcm5\times2\farcm5$) likely covers most of, if not all of, 
the entire CSE of CIT 6 in the $^{12}$CO $J$\,=\,2--1 and $^{13}$CO 
$J$\,=\,2--1 line emission. 

We define the systemic velocity ($\vsys=-2\,\kmps$) as the middle of two 
peaks of the $^{13}$CO $J$\,=\,2--1 spectrum from the central $1\arcsec$ 
region, presented in (a) panel. The velocity $v_z$ hereafter represents 
the velocity with respect to the systemic velocity ($v_z=\vLSR-\vsys$). 
The $^{13}$CO spectrum in the central $1\arcsec$ region is sharply doubly 
peaked at $v_z=\pm16\,\kmps$, beyond which the intensity levels quickly 
drop to zero within 2\,\kmps. 
The outer parts of $^{13}$CO emission seem to have an extra velocity 
component around $-10\,\kmps$, as seen in (b) panel, in a comparable 
brightness temperature level with that of the $-16\,\kmps$ component. 
In (c) and (d) panels, this component predominates over the $-16\,\kmps$ 
component that appears as a broader shelf in the blueshifted velocity 
range. In contrast, the red peak velocity does not change much with the 
integration radius. 

The $^{12}$CO $J$\,=\,2--1 spectral asymmetry depends on the integration 
area. In Figure\,\ref{fig:spe}(a), the central region ($r\leq1\arcsec$) has 
a red-skewed doubly peaked spectrum with the redshifted peak (at $v_z=14$%
\,\kmps) about 30\% brighter than the blueshifted peak (at $-8\,\kmps$). 
In order to quantify the spectral profile asymmetries varying with the 
integration area, we define the fraction of asymmetry as the emission 
excess of the positive velocity component (at 0 to 16\,\kmps) beyond 
the negative velocity component (at $-16$ to 0\,\kmps). In the central 
1\arcsec\ region shown in (a), the positive velocity emission is about 
40\% brighter than the negative velocity emission, i.e., the fraction 
of asymmetry is 40\%. As the integration area increases, the fraction 
of asymmetry is reduced to 30\% in the (nearly) flat-topped profile in 
$r\le5\arcsec$ region shown in (b), and to 10\% and 5\% in (c) and (d), 
respectively. This decrease of the fraction of asymmetry with the 
integration radius implies that the source of spectral asymmetry is placed 
in the central region of the $^{12}$CO line emission. For comparison, 
the fractions of asymmetry of the $^{13}$CO $J$\,=\,2--1 line spectra 
are within $\pm\,5\%$ regardless of integration area. 

\subsection{Averaged Radial Distribution}\label{sec:prf}

Considering the CSE of an AGB star having radial flows with a nearly 
constant velocity, the channel at the expansion velocity exhibits emission 
integrated along the line of sight. Hence, it is subject to an optically 
thick condition. 
On the other hand, the optical depth at the systemic velocity channel, 
exhibiting the emission from the plane of the sky, is relatively small. 
Therefore, in Figure\,\ref{fig:pro}, we use the systemic velocity channel 
to compare the radial profiles of the $^{12}$CO $J$\,=\,2--1 and $^{13}$CO 
$J$\,=\,2--1 brightness temperatures averaged over azimuthal angles. 

In order to examine the brightness distribution from small to large 
scales, we generate maps for four different combinations of SMA and 
SMT data: SMA-only (yellow; beam size $\sim0\farcs5$), joint of SMA 
and SMT (black; $\sim2\arcsec$), joint of SMA subcompact and compact 
configuration with SMT (blue; $\sim6\arcsec$), and SMT-only (red; 
$\sim33\arcsec$). All four curves for $^{12}$CO $J$\,=\,2--1 are 
consistent with each other (except for the SMA-only profile beyond 
10\arcsec; see next paragraph), following a monotonically decreasing 
curve from $r\sim0\farcs2$ to 60\arcsec. For the $\sim20$ times weaker 
$^{13}$CO $J$\,=\,2--1 emission, the trend can be found by connecting 
the profiles from different configurations at sufficient signal-to-noise. 
The $^{12}$CO profile changes slope (in logarithmic scale) smoothly from 
$\sim-0.7$ for the inner part ($r\la5\arcsec$) to $\sim-2.0$ beyond 5\arcsec. 
On the other hand, the $^{13}$CO profile keeps the shallow slope $\sim-0.7$ 
up to the radius $\sim20\arcsec$ and then drops with a markedly steeper slope 
$\sim-3.0$. As a result, the $^{12}$CO and $^{13}$CO $J$\,=\,2--1 line ratio 
is $\sim30$ in the inner 5\arcsec\ region, $\sim10$ at $\sim20\arcsec$, and 
$\sim20$ beyond 30\arcsec.

The only significant deviation from the rest configuration data occurs in 
the SMA-only profile, dramatically dropping beyond its maximum recoverable 
scale ($r\sim10\arcsec$) due to the finite minimum projected baseline 
($\sim5$\,m). Despite such missing flux in the SMA interferometry beyond 
the maximum recoverable scale, the minimum baseline of the SMA-only data 
was sufficiently short so that the resulting brightness temperature reduction 
($<2$\,K) does not significantly change the profile within $\sim10\arcsec$. 
Indeed, the convolution of the SMA-only map with a $23\arcsec$-sized Gaussian 
beam yields the peak and integrated brightness temperatures of 3.74\,K (at 
$v_z=0\,\kmps$) and 98.0\,K\,\kmps, essentially identical with the values from 
a 23\arcsec-beam single-dish observation \citep[3.72\,K and 97.3\,K\,\kmps,]
[]{olo93}. The SMA-related profiles (yellow, black, and blue) have large error 
bars beyond the SMA primary beam radius $\sim30\arcsec$.

\section{FINE STRUCTURE}\label{sec:fin}
\subsection{Spiral-shell Pattern}\label{sec:spi}

Figure\,\ref{fig:chm} displays the $^{12}$CO $J$\,=\,2--1 channel 
map of CIT 6 from the SMA observations with a $\sim0\farcs5$ resolution for 
a channel velocity width of 5\,\kmps\ (in pseudo blue color). For comparison, 
it is overlaid by the HC$_3$N $J$\,=\,4--3 map (in pseudo red color) taken 
with the JVLA \citep{cla11}. The broken-shells of the HC$_3$N emission for 
CIT 6 were used in the previous spiral-shell modeling by \citet{kim13}. The 
central velocity channels (middle row panels) indeed reveal spiral patterns 
in $^{12}$CO, and the patterns at higher velocities (top and bottom panels) 
are rather ring-like. Both of these features agree well with the spiral-shell 
geometry \citep[see e.g., Figure\,8 of][]{kim12b} implying the existence of 
the companion to CIT 6 and hence diagnostics for the orbital motion of the 
binary system \citep{kim13}. 

A single spiral pattern is particularly discerned in the $v_z=5\,\kmps$ 
channel (middle right panel of Figure\,\ref{fig:chm}). The pattern has a 
shape rolling inward in the clockwise sense, indicating the orientation of  
binary orbital motion. Inner two windings are apparent in the $^{12}$CO map 
with the brightness level over $20\sigma$. On the other hand, the pattern 
beyond $\sim5\arcsec$ is rather sporadic. The arm at $\sim8\arcsec$ to the 
west is perceptible at $\sim6\sigma$ with a higher arm-interarm contrast 
than in other directions, caused by a reduction of the interarm brightness 
level (see below). 

Figure\,\ref{fig:prf} displays the radial profiles of the $^{12}$CO line 
emission at the systemic velocity channel ($v_z=0\,\kmps$). 
Each colored curve is defined along the noted PA by azimuthal average with 
the width of 20\arcdeg, and the bottom profile is averaged over 360\arcdeg. 
The arms are partially resolved with the SMA beam size $\sim0\farcs5$, 
which would form otherwise almost monotonically decreasing curves (as in 
Figure\,\ref{fig:pro}, the profile for the SMA--SMT combined image with a 
larger $\sim2\arcsec$-sized beam). The radii of local peak positions differ 
with PA, indicating the non-spherical nature as expected in a spiral model. 
The emission in the spiral arm appears as 20\% fluctuation of total emission, 
and the peak values in the arm regions exceed the adjacent interarm values 
by a factor of $\sim1.5$ to $\sim3$. 

Except for the west side (PA of 270\arcdeg), no systematic difference 
is found in the shape of directional profiles, unlike in close binary 
systems presenting the envelope flattening onto the orbital plane 
\citep[see][]{hug09,mau13}. 
The brightnesses decrease to $3\sigma$ (color-coded dotted lines in 
Figure\,\ref{fig:prf}) at $r\sim9\arcsec$ except for the profile to 
the west, where the brightness reaches $3\sigma$ at a smaller radius 
$r\sim5\arcsec$. The overall profile to the west is 25\% steeper than 
in all the other directions. This deficiency of the $^{12}$CO line 
emission to the west probably indicates a lower mass ejection along 
this direction due to, e.g., anisotropic dust formation near the stellar 
surface, or non-constant orbital speed of the star. 

The position-velocity (P-V) diagrams\footnote{\citet{kim13} suggested 
the PA of 10\arcdeg\ as the intersection between the plane of the sky 
with the orbital plane of the binary stars, along which the P-V diagram 
is the most symmetric and perpendicular to which (PA of 280\arcdeg) the 
P-V diagram is the most asymmetric.} in Figure\,\ref{fig:pvd}(a)--(b) 
better exhibit the pattern in a continuous and definitive shape. The 
arm patterns in Figure\,\ref{fig:pvd}(a) are rather regularly arranged 
as being approximated as parabolas with the maximum offsets of 2\farcs5, 
4\farcs0, 5\farcs7, and 8\farcs3 for the PA of 10\arcdeg\ (positive 
offset in the panel) with intervals of 1\farcs5, 1\farcs7, and 2\farcs6; 
for the PA of 190\arcdeg\ (negative offset), the maximum offsets 
of these parabolas are 2\farcs0, 3\farcs5, 5\farcs0, 6\farcs7, and 
9\farcs3 with intervals of 1\farcs5, 1\farcs5, 1\farcs7, and 2\farcs6. 
The observed P-V diagram patterns tend to have an arm interval of 
$\sim1\farcs5$ in the inner $\sim7\arcsec$ region, but have a larger 
interval of 2\farcs5 between the observed outermost arms located at 
$\sim6\arcsec$--9\arcsec. This change of arm interval can be due to 
a temporal decrease of the wind velocity. 

\subsection{Emission Peak Position with Channel Velocity}\label{sec:max}

Upon a close inspection of the zoomed-in channel maps for the $^{12}$CO 
emission (Figure\,\ref{fig:cpk}), the shape of emitting region is not 
circular and varies with channels. We further investigate the positional 
change of channel peaks by fitting the observed offsets along the axes 
of RA and Dec with sinusoidal functions in the form of $f=a_0+a_1\sin
(2\pi(v_z-a_3)/a_2)$, as shown in Figure\,\ref{fig:fit}(a)--(b). 
The parameter $a_0$ represents the average emission peak position, $a_1$ 
the scale of maximum positional shift, and $a_2$ the channel velocity scale 
for which the positional variation of emission peaks occurs. Here, we focus 
on the velocity range between $-13$ and $16\,\kmps$. The rapid change of 
peak positions at the high-end velocities (related to a bipolar outflow) 
is discussed in Section\,\ref{sec:jet}.

These separately fitted sinusoidal functions for RA and Dec both have 
the same $a_2$ parameter of 17.7\,\kmps, indicating that the emission 
peak position varies periodically in the scale of the expansion 
velocity of the envelope. Because of the consistency of cycles in RA 
and Dec, the combined fitting results lead to a closed ellipse as 
a function of channel velocity as shown in Figure\,\ref{fig:fit}(c). 
It approximates an ellipse with an ellipticity $\sim0.75$ and a major 
axis on the scale $\sim0\farcs15$ with the PA $\sim-40\arcdeg$. 
The peak positions of channels at $\sim-8\,\kmps$ and $\sim10\,\kmps$ 
become closer to the continuum peak position, which we defined as the 
coordinate center. In average, the channel peaks of the $^{12}$CO 
$J$\,=\,2--1 emission are located at the northwest side of the 
continuum peak position ($<\Delta\,\rm RA>=-0\farcs07\pm0\farcs06$; 
$<\Delta\,\rm Dec>=0\farcs04\pm0\farcs07$). 
The rotation of peak positions along the channels were also reported 
by \citet{lin00} for the HCN $J$\,=\,1--0, HNC $J$\,=\,1--0, HC$_3$N 
$J$\,=\,10--9, and SiS $J$\,=\,5--4 line emission over larger scales 
($\sim0\farcs5$--2\arcsec); the loci of their channel peaks could be 
connected to a curve elongated in the northwest-southeast direction 
across the line profile, similar to the PA of ellipse in our fitting 
result of the $^{12}$CO peak positions. 

The positional measurement uncertainty is given by the largest value 
amongst (1) the beam size divided by the peak signal-to-noise ratio 
($\la0\farcs01$ in most channels; it increases to $\sim0\farcs06$ at the 
line edge channel), (2) the positional (or phase) uncertainty of the 
gain calibrators derived by the same method experiments ($<0\farcs03$), 
and (3) the image pixel size (0\farcs03). However, these provide a 
lower limit of the positional uncertainty as the $^{12}$CO emission 
is extended. A firm conclusion on the spatial change of the emission 
peak (in a rotating sense) with velocity and its relationship with the 
binary motion in an inclined orbital plane are the subject of higher 
spatial and spectroscopic resolution observations.

\subsection{Bipolar Outflow}\label{sec:jet}

The $^{12}$CO spectrum in Figure\,\ref{fig:spe}(a) for the central 1\arcsec\ 
region exhibits a small bump at $v_z\sim-19\,\kmps$, distinguished from the 
main line by a valley at $v_z=-16\,\kmps$. Such a feature is also found in 
P-V diagrams, appearing as an island located at $-21\la v_z\la-17\,\kmps$ 
(see Figure\,\ref{fig:pvd}(a)--(b)). This component contributes about 1.4\% 
in the total flux. 

Other interferometric observations showed the blue bump feature 
in several pPNe \citep[e.g., IK Tau, $\chi$ Cyg, IRAS 20028$+$3910, and 
IRAS 23321$+$6545 in Figures\,4, 22, and 23 of][]{cas10}, where fast 
bipolar outflow structures are well developed. These spectral features were 
interpreted as an indication of heating due to shock interaction between 
the inner fast collimated winds and the outer slow nearly spherical winds 
ejected during the AGB phase. The emission from the hot shocked region is 
vulnerable to self-absorption by the surrounding cooler medium. The blue 
bump feature, in addition to the depressed blue peak below the red peak in 
Figure\,\ref{fig:spe}, can be also induced in a perfectly spherical wind 
due to partial self-absorption of the central warmer emission by the outer 
cooler envelope \citep{mor85}. 
Self-absorption was introduced by both \citet{mor85} and \citet{cas10} in 
explaining the blue bump features, but the geometry and physical mechanism 
of the hot inner source are different from each other. 

As clearly seen in Figure\,\ref{fig:jet}, the observed blue bump of 
CIT 6 originates from a bipolar outflow. Figure\,\ref{fig:jet} is the 
same as the highest velocity channels in Figure\,\ref{fig:chm}, but in 
blue and red contours for the blueshifted ($-19\,\kmps$) and redshifted 
(19\,\kmps) components, respectively. For comparison, the emission peak 
at the systemic velocity is also plotted in green contours, showing its 
position significantly close to the continuum peak position (gray-scaled 
image). In contrast, the high-velocity emission peaks are definitely offset 
from the continuum peak position to the northeast-southwest direction. With 
the directional similarity, we consider these high velocity components as 
the gaseous counterpart of the near-infrared bipolar nebula \citep{sch02}. 

The $-19\,\kmps$ component has an elongated shape to the northeast. It 
is distributed in a conical shape about the PA of 55\arcdeg, which is 
consistent with the PA of the near-infrared reflection nebula (0\arcdeg%
--80\arcdeg) shown in \citet{sch02}. The size ($\sim3\arcsec$ at $3\sigma$) 
of this component is comparable to the size of the near-infrared bipolar 
nebula. Its conical shape has the vertex close to, but slightly south of, 
the continuum peak position. Assuming that the continuum peak position 
coincides with the location of the visible red point source (the probable 
carbon star) in \citet{mon00}, the location of the visible blue point source 
(possibly the companion, $\la0\farcs2$ south of the red source) is marked by 
a cross symbol in Figure\,\ref{fig:jet}. With the current angular resolution, 
it is difficult to distinguish whether the conical outflow originates from 
the carbon star or the companion. 

The redshifted component is considerably smaller ($\sim1.5\arcsec$ at 
$8\sigma$) than the blueshifted component. The redshifted component at 
$\sim19\,\kmps$ extends to the south or southeast in Figure\,\ref{fig:jet}. 
The central spectrum in Figure\,\ref{fig:spe}(a) does not show any special 
feature at $\sim19\,\kmps$, unlike the blueshifted component appears as 
a small bump. However, Figure\,\ref{fig:cpk} and \ref{fig:fit} show that 
the position of emission peak in each channel beyond 16\,\kmps\ gradually 
shifts to the southeast as the velocity increases. This shift is noticeable 
because the emission peak of each channel remains within the central 0\farcs2 
regime in most of velocity range ($-13$ to 16\,\kmps). 
The 19\,\kmps\ component appears to consist of two features, one in 
a PA of 190\arcdeg\ and the other in 150\arcdeg. This is most evident for 
the red contour corresponding to $12\sigma$. The component along the PA of 
190\arcdeg\ is approximately in the direction opposite to the blue component. 
The presence of red component along the PA of 150\arcdeg\ implies multipolar 
nature, together with a discrete blue component at $8\sigma$ level in the 
opposite (northwest) direction.

We note a cometary-shaped silicate feature at 9.7\,$\mu$m \citep{lag05} 
along the PA of $\sim230\arcdeg$ with a large extension up to 7\arcsec\ 
(at 1\% of the peak intensity), which is exactly on the opposite side 
of our $-19\,\kmps$ component. 
Currently, however, there is no strong evidence to further relate the 
9.7\,$\mu$m feature to the redshifted component of the bipolar outflow.

\section{DISCUSSION}\label{sec:dis}
\subsection{One-Dimensional Radiative Transfer Model}\label{sec:rtm}

\begin{figure*} 
  \plotone{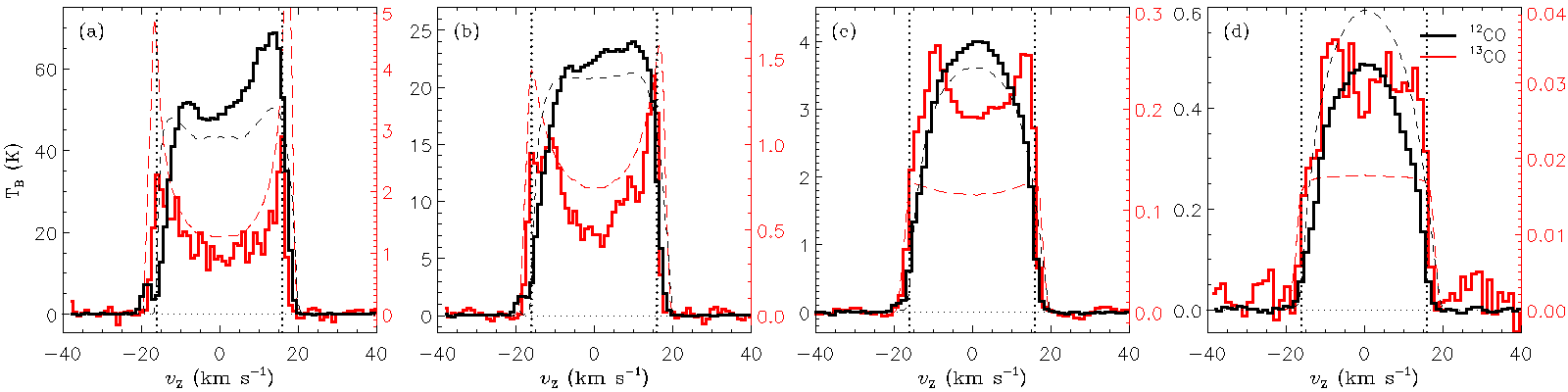}
  \caption{\label{fig:spe}
    Spectra of the observed $^{12}$CO $J$\,=\,2--1 (black solid line) 
    and $^{13}$CO $J$\,=\,2--1 (red solid) emission from (a) the central 
    region ($r\leq1\arcsec$) of the SMA-SMT combined images, (b) the 
    $r\leq5\arcsec$ region, (c) the single pointing with the SMT (beam 
    size $\sim33\arcsec$), and (d) the entire region of the SMT grid-map 
    ($\sim2\farcm5\times2\farcm5$) in units of brightness temperature. 
    One-dimensional radiative transfer model spectra with the $^{12}$CO%
    /$^{13}$CO abundance ratio of 50 is presented by dashed lines in the 
    same color-codes. The vertical axes of $^{12}$CO line spectra are 
    scaled by 15 from the $^{13}$CO axes. Vertical dotted lines mark 
    $v_z=\pm16\,\kmps$, where $v_z=\vLSR-\vsys$ and $\vsys=-2\,\kmps$. 
  }
\end{figure*}

\begin{figure*} 
  \epsscale{1.1}
  \plottwo{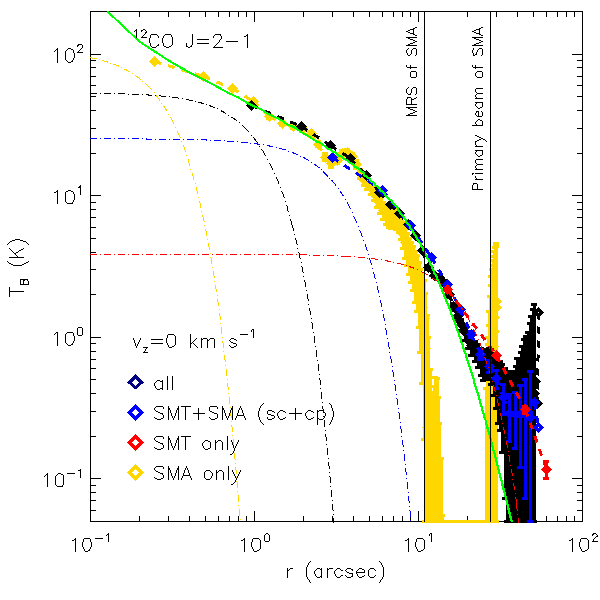}{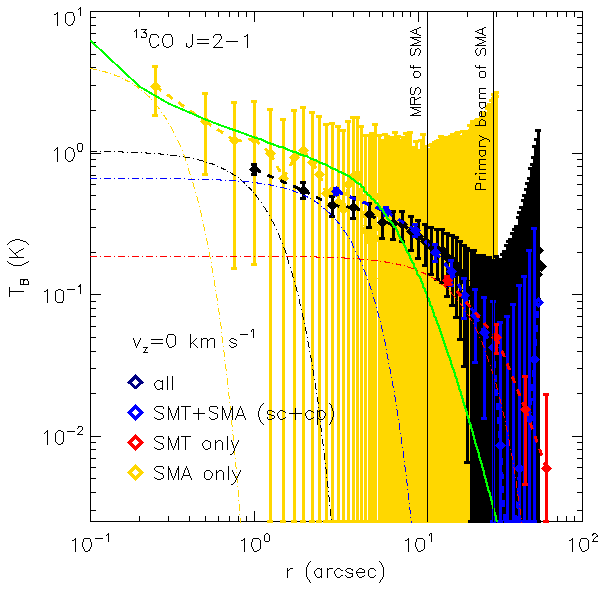}
  \caption{\label{fig:pro}
    Radial profiles of $^{12}$CO $J$\,=\,2--1 (left panel) and $^{13}$CO 
    $J$\,=\,2--1 (right panel) at the systemic velocity (width of 5\,\kmps) 
    in different configurations of the array. Our radiative transfer model 
    is overlaid by green curves (see text for detail). Dash-dotted lines 
    represent the restored Gaussian beams in the same color codes for the 
    data points. The maximum recoverable scale and primary beam of the SMA 
    are marked in each panel.
  }
  \epsscale{1}
\end{figure*}

\begin{figure*} 
  \plotone{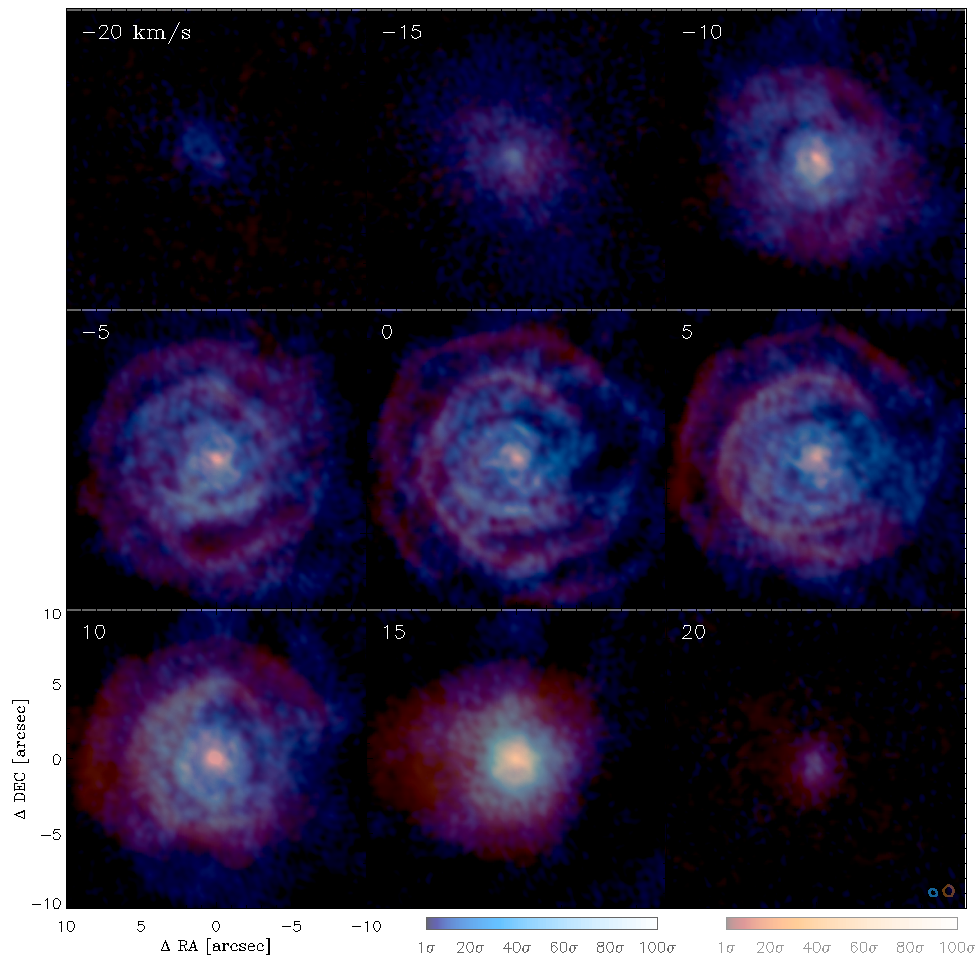}
  \caption{\label{fig:chm}
    Channel map of the $^{12}$CO $J$\,=\,2--1 emission (blue) from the SMA 
    observation, overlaid with that of HC$_3$N $J$\,=\,4--3 emission (red; 
    \citealp{cla11}). Each panel displays the channel image averaged over 
    5\,\kmps, and is labeled by $v_z=\vLSR-\vsys$ (that is, the channel 
    velocity with respect to the systemic velocity $\vsys=-2\,\kmps$). The 
    image center is set to the position of the peak of continuum emission 
    at 226.2\,GHz. The synthesized beam sizes of the $^{12}$CO and HC$_3$N 
    lines are denoted at the bottom right corner in the corresponding colors. 
    Colorbars are marked in units of the noise levels for a velocity width 
    of 5\,\kmps; $\sigma=12\,\mJpb$ (1.1\,K) for the $^{12}$CO map, and 
    $\sigma=0.24\,\mJpb$ (0.46\,K) for the HC$_3$N map. 
  }
\end{figure*}

We have performed simple, one-dimensional, non-local thermodynamic 
equilibrium radiative transfer calculations using the SPARX code%
\footnote{SPARX (Simulation Package for Astrophysical Radiative 
Transfer) is a software package for calculating molecular excitation 
and radiative transfer of dust continuum and molecular lines at 
millimeter/submillimeter wavelengths (Liu et al., in prep.).} to 
understand the overall distribution of the $^{12}$CO and $^{13}$CO 
molecules in the CSE of CIT 6. 
Since our focus here is on the overall CSE distribution, the observed 
fine structures (as described in Section\,\ref{sec:fin}) are ignored 
in the modeling; see \citet{kim13} for a spiral-shell model for 
CIT 6, characterizing the HC$_3$N $J$\,=\,4--3 pattern in channel maps. 
In Figures\,\ref{fig:spe} and \ref{fig:pro}, we present a simple model for 
an AGB envelope expanding with a constant expansion velocity $V_w=18\,\kmps$, 
a mass loss rate $\dot{M}=8\times10^{-6}\,\Mspy$, a temperature distribution 
$T(r)=2000\rm\,K\,(r/4\,AU)^{-0.7}$, a stochastic velocity of 1\,\kmps, 
constant fractional abundances of molecules ($f_{\rm^{12}CO}=1\times10^{-3}$ 
and $f_{\rm^{13}CO}=2\times10^{-5}$), and a distance of 400\,pc. 

This model reproduces reasonably well the average radial profile of the 
observed $^{12}$CO $J$\,=\,2--1 (Figure\,\ref{fig:pro}, green curve) and 
its spectra in different length scales (Figure\,\ref{fig:spe}, dashed 
line), providing some insights to physical conditions of the overall CSE 
of CIT 6. 
The optical depth of the $^{12}$CO $J$\,=\,2--1 model at the systemic 
velocity is low ($\sim0.5$) at the center, increases to $\sim2$ at 
$r\sim10\arcsec$, and decreases back to $\sim0.5$ at $r\sim40\arcsec$. 
On the other hand, the expansion velocity channel has the large optical 
depth greater than 4 for almost all emitting regions with the peak value 
over 10 at $r\sim4\arcsec$. 
The over-predicted flux in Figure\,\ref{fig:spe}(d) may indicate that 
the $^{12}$CO abundance in the outermost region is lower than in the 
inner region possibly because of photodissociation due to interstellar 
ultraviolet photons. This interpretation is consistent with the radius 
at which the abundance drops to half of its photospheric value derived 
to be $\sim30\arcsec$ by e.g., \citet{sch01}, based on \citet{mam88}. 

The radial profile of $^{13}$CO $J$\,=\,2--1, however, is not achieved by 
merely changing the \emph{single} molecular fractional abundance with the 
same fixed mass loss rate and temperature profile as in $^{12}$CO model. 
The $^{13}$CO modeled and observed spectra are not too far off in the 
inner part (Figure\,\ref{fig:spe}(a)--(b)); but the outer envelope of 
CIT 6 contains significantly more $^{13}$CO emission compared to the model 
(Figure\,\ref{fig:spe}(c)--(d)). In fact, as seen in Figure\,\ref{fig:pro}, 
the $^{13}$CO radial profile of CIT 6 is relatively flatter and more 
attributed by extended component than in $^{12}$CO. The significant 
difference in slopes of the $^{12}$CO and $^{13}$CO radial profiles 
may imply a variation of their abundance ratio with radius. 

The $^{12}$CO/$^{13}$CO abundance ratio 50 in the presented model is somewhat 
larger than the value $\la35$ in the literature \citep[and reference therein]
{mil09}. Our choice for the $^{13}$CO fractional abundance results from 
matching the brightness temperatures in the inner region closely the observed 
values, at the expense of the outer region. We have checked that the modeled 
$^{13}$CO spectra with a ratio $\sim30$ can reproduce the SMT single-dish 
observational data (i.e., for Figure\,\ref{fig:spe}(c)--(d)), consistent with 
the earlier models by e.g., \citet{mil09}, but resulting in a mismatch in the 
inner envelope. 
It may indicate that $^{12}$CO/$^{13}$CO abundance ratio varies with radius. 

\subsection{Comparison with HC$_3$N}\label{sec:cmp}

Considering both the $^{12}$CO $J$\,=\,2--1 and HC$_3$N $J$\,=\,4--3 
high resolution maps together, four to five windings of the spiral-shell 
pattern are found. In terms of their location, there are close agreements 
between the $^{12}$CO and HC$_3$N patterns manifested in the channel maps 
(Figure\,\ref{fig:chm}) and the P-V diagrams (Figure\,\ref{fig:pvd}). For 
example, to the west where the HC$_3$N pattern is broken, the $^{12}$CO 
pattern is present and smoothly connects the HC$_3$N segments from the 
north to the south. Figure\,\ref{fig:pvd}(a) and (c) show the $^{12}$CO 
and HC$_3$N patterns along the PA of 10\arcdeg, and their overlay in 
Figure\,\ref{fig:pvd}(e) demonstrates their coincidence. The local peak 
positions are marked by open circles (blue for $^{12}$CO; red for HC$_3$N), 
illustrating their good match in the P-V diagram. In the bottom panels of 
Figure\,\ref{fig:pvd} along the PA of 280\arcdeg, the HC$_3$N pattern in 
(d) is prominent in the outer shells, while the $^{12}$CO emission in (b) 
is centralized and the outer pattern fading away. However, their overlay 
in (f) again shows a good correlation. 

Albeit there is a good correlation in the pattern locations, some 
differences can be found between the $^{12}$CO $J$\,=\,2--1 and 
HC$_3$N $J$\,=\,4--3 maps. The $^{12}$CO emission is omnipresent 
in the CSE of CIT 6, while the HC$_3$N line emission is absent 
toward the center and west sides (see Figure\,\ref{fig:chm}, e.g., 
the panels for $v_z=0\,\kmps$ and 5\,\kmps). The $^{12}$CO map 
exclusively reveals the first and second windings in the central 
5\arcsec\ region, while their HC$_3$N counterparts are missed. The 
P-V diagrams (Figure\,\ref{fig:pvd}(a)--(b)) further exhibit the 
richness of the $^{12}$CO structure within the central 5\arcsec, 
where HC$_3$N is lacking as shown in Figure\,\ref{fig:pvd}(c)--(d). 
The second and third $^{12}$CO windings appear broken in the 
HC$_3$N map in the west side. The $^{12}$CO structures appearing 
at $\sim5\arcsec$ and $\sim10\arcsec$ in Figure\,\ref{fig:pvd}(b) 
are completely missed in the corresponding P-V diagram of HC$_3$N, 
i.e., Figure\,\ref{fig:pvd}(d). 

To the west side where the HC$_3$N $J$\,=\,4--3 emission was absent, 
the $^{12}$CO $J$\,=\,2--1 arm pattern appears but its radial profile 
is steeper than in any other directions. It implies that the west side 
is indeed special in both molecules, probably indicating anisotropic 
mass loss. Nevertheless, the complete non-detection of the western part 
of spiral arm pattern in  HC$_3$N $J$\,=\,4--3 implies that the chemical 
effects on HC$_3$N may be not negligible. 

The $^{12}$CO emission is seen in both the compressed spiral arm and 
the interarm region, whereas the HC$_3$N emission seems to be mostly 
enhanced in the outer edge of the shock front as following. In the 
outer region where the HC$_3$N pattern is prominent ($\sim3\arcsec$ 
to 10\arcsec), the spiral-shell pattern appears better in HC$_3$N with 
high arm-interarm contrasts. The HC$_3$N pattern segments are narrow 
(comparable to its synthesized beam size $\sim0\farcs7$), while the 
$^{12}$CO arm pattern is broader and the arm-interarm contrasts are 
smaller. In addition, the HC$_3$N emission peaks marked by red circles 
in Figure\,\ref{fig:pvd}(e)--(f) tend to be located slightly outside of 
the $^{12}$CO emission peaks (blue). All these features may indicate 
that HC$_3$N is sufficiently excited at the edge of the spiral arm 
region tracing the shock front. 

The critical density for a molecular line transition, at which the emission 
induced by molecular collisions dominates over spontaneous emission, gives 
an approximate lower limit on the gas density in the emitting region of 
such line transition. The critical density of $^{12}$CO $J$\,=\,2--1 is 
nearly independent of temperature ($\sim10^4\rm\,cm^{-3}$), while that of 
HC$_3$N $J$\,=\,4--3 is higher ($>2.5\times10^4\rm\,cm^{-3}$ for $<100$\,K) 
and drops with temperature\footnote{The critical densities are calculated 
based on the Leiden Atomic and Molecular Database \citep{sch05} with a 
simple two-level balance.}. It implies that the HC$_3$N $J$\,=\,4--3 is 
a higher density tracer at the typical average temperatures of AGB CSEs 
($<100$\,K). In the radius range ($r\sim3\arcsec$\,--\,10\arcsec) where 
the HC$_3$N pattern of CIT 6 is observed, a low interarm number density 
$\la2\times10^3$\,cm$^{-3}$ and gas temperature $\la40$\,K are expected 
\citep{din09,kim13}. Therefore the lack of HC$_3$N line emission in the 
interarm region is reasonable. Within the spiral arm, a number density of 
$10^4$--$10^6$\,cm$^{-3}$ with a temperature of several hundreds Kelvins 
is possible \citep{kim13}. Hence, molecular collisions are sufficient for 
exciting HC$_3$N $J$\,=\,4--3, explaining the high arm-interarm contrast 
in the HC$_3$N map. On the other hand, the critical density of $^{12}$CO 
$J$\,=\,2--1 is independent of gas temperature, making this emission well 
excited in both arm and interarm regions, and reducing the contrast between 
the arm and interarm regions. 

\subsection{Signatures of Eccentric Orbit}\label{sec:ecc}

The systemic velocity channel interestingly shows a double spiral feature in 
the central $5\arcsec$ region, as shown in Figure\,\ref{fig:dsp} (see 
also the middle panel of Figure\,\ref{fig:chm} for a larger field), which was 
not generated in the previous hydrodynamic simulation assuming a constant 
wind velocity of 17\,\kmps\ \citep{kim13}. A slower wind at the locus of the 
companion may facilitate generation of another spiral \citep[see e.g., 
Figure 3 of][]{moh12} provided that the flow velocity is less than the 
escape velocity from the companion's gravitational potential. The inner 
double spirals are expected to merge together within one to two windings 
because of their different pattern propagation speeds. This merging seems 
to occur at around 4\arcsec\ in CIT 6, leaving the one-armed spiral 
pattern in the outer part of Figure\,\ref{fig:chm} (middle panel) as traced 
in the HC$_3$N and CO line emissions. 

However, the projected binary separation of $\la80$\,AU \citep{mon00} 
is far beyond several stellar radii, where dust grains form and stellar 
radiation pressures on them accelerate the CSE flow to the terminal 
velocity \citep{hof07}. We speculate that a highly eccentric orbit may 
allow the companion to pass the acceleration zone, possibly forming a 
double spiral in the inner part of the CSE of CIT 6. An orbit with the 
periastron closer than the acceleration zone radius (assuming 20\,AU) 
and the apocenter at $r\ga80$\,AU requires an eccentricity $\ga0.6$.

For an eccentric orbit, the orbital speed of a mass-losing star is fastest 
at the periastron. Therefore the accumulated mass ejection to this direction 
is lower than in other directions. The lack of $^{12}$CO interarm emission 
to the west (along a PA of 270\arcdeg\ in Figure\,\ref{fig:prf}), closely 
resembles the column density map of a highly eccentric orbit model in \citet
[see e.g., the model with an eccentricity of 0.8]{he07}. The anisotropic mass 
loss due to the eccentric orbit alters the gas density particularly in the 
interarm region, which is similar to our finding from the $^{12}$CO brightness 
radial profiles. In this case, the periastron would be on the west side. 

\citet{cer15} further employed a concept of \emph{periodic} mass loss 
enhancement triggered by a \emph{close} fly-by companion in their piston 
model, creating a more complex pattern on the CSE. If this strong tidal 
interaction model applies to CIT 6, the periastron, at which a larger 
mass loss could occur, would be on the opposite, east side leading to a 
higher average intensity. This situation is less likely for a wide binary 
like CIT 6, although the necessary binary separation for the strong tidal 
interaction between the stars is not well studied.

\section{SUMMARY}\label{sec:sum}

We have observed CIT 6 in all possible (subcompact, compact, extended, and 
very extended) array configurations of the SMA from 2013 January to 2014 
March, and combined with the SMT mapping observations on 2015 January. 
Our findings from the $^{12}$CO $J$\,=\,2--1, $^{13}$CO $J$\,=\,2--1, and 
continuum emission are summarized as follows:

\begin{itemize}
  \item CIT 6's continuum position at the epoch 2013.1 is (RA, Dec)$=$%
    (10$^{\rm h}$16$^{\rm m}$02\fs259, $+$30\arcdeg34\arcmin19\farcs18) 
    in J2000, updating the proper motion measurement to be ($-17\pm4$, 
    $16\pm4$)\,mas\,yr$^{-1}$.
    On the other hand, the $^{12}$CO $J$\,=\,2--1 emission peak position 
    rotates counterclockwise with the channel velocity, staying on the 
    northwest side ($<\Delta\,\rm RA>=-70\pm60$\,mas; $<\Delta\,\rm 
    Dec>=40\pm70$\,mas) of the dust continuum emission center. The positional 
    shift of molecular line emission peak in the proper motion direction 
    is noticeable. 
  \item Continuum emission at 226.2\,GHz from the SMA observations best 
    fits to a two-component model consisting of an unresolved source (of 
    flux 26.4\,mJy) and an extended component with a size $\la4\arcsec$ 
    (34.5\,mJy). The unresolved continuum source is also observed at 
    350.6\,GHz and has a spectral index $\sim-2$, consistent with thermal 
    blackbody emission from a radio photosphere with a diameter of 8\,AU 
    and the temperature of 2000\,K. 
  \item The systemic LSR velocity of $-2\,\kmps$ is defined as the 
    middle velocity between the double peak spectral profile of the 
    $^{13}$CO $J$\,=\,2--1 line in the innermost 1\arcsec\ region. 
    With respect to the systemic velocity, the peaks are located 
    at $\pm16\,\kmps$ and its brightness temperature reaches zero 
    at $\sim\pm18\,\kmps$. The outer envelope has an additional 
    $-10\,\kmps$ component, brighter than the $-16\,\kmps$ peak.
  \item The $^{12}$CO $J$\,=\,2--1 emission spectral profile varies 
    with integration radius; a red-skewed double peak profile (with 
    a small blue-end bump) over $r\le1\arcsec$, a nearly flat-topped 
    profile over $r\le5\arcsec$, and a parabolic shaped profile from 
    the single dish observations with a $\sim33\arcsec$-sized beam. 
    It shows the optical depth change with radius; i.e., the innermost 
    envelope is relatively optically thin and a significant fraction 
    of 5\arcsec\ region becomes optically thick. 
  \item The spatially-averaged radial and spectral profiles of $^{12}$CO 
    $J$=2--1 in all length scales are reasonably well reproduced by a simple 
    spherical radiative transfer model. However, the fixed fractional 
    abundance model fails in matching the overall radial distribution of 
    the observed $^{13}$CO $J$=2--1 emission. This may suggest that the 
    $^{12}$CO/$^{13}$CO abundance ratio varies in the CSE of CIT 6 from 
    $\sim30$ in the outer envelope to $\sim50$ in the inner envelope. 
  \item The high resolution $^{12}$CO $J$\,=\,2--1 map (from the SMA) 
    reveals a spiral-shell pattern appearing as a 20\% fluctuation of 
    the average profile, due to the suggested binary orbital motion. 
    This pattern is undetected in the $^{13}$CO $J$\,=\,2--1 emission 
    likely due to sensitivity, as 20\% of the observed $^{13}$CO 
    maximum brightness corresponds to $\sim1\sigma$ only. 
  \item The previously observed, arc-shaped pattern in the HC$_3$N 
    $J$\,=\,4--3 emission map turns out to be an incomplete form of 
    the spiral-shell pattern appearing in the $^{12}$CO $J$\,=\,2--1 
    map. The HC$_3$N segments (broken to the west side and central 
    3\arcsec\ region) are connected by the $^{12}$CO pattern. 
    The HC$_3$N $J$\,=\,4--3 line emission likely traces the shocked 
    arm region with high arm-interarm contrasts.
  \item A bipolar outflow of CIT 6 is found for the first time in the 
    $^{12}$CO $J$\,=\,2--1 channel map. The blue component (at $-19\,
    \kmps$) is separated from the AGB envelope by a gap at $-16\,\kmps$, 
    clearly seen in the P-V diagrams. The red counterpart (at 19\,\kmps) 
    is smaller than the blue component of the bipolar outflow, and points 
    to a nearly perpendicular direction. 
  \item In the zoomed-in channel map and spectrum, the $\sim10\,\kmps$ 
    channels exhibit the strongest emission, indicating that the spectral 
    asymmetry of the $^{12}$CO $J$\,=\,2--1 emission originates from the 
    innermost region. In contrast, the source of the red-skewed profile 
    of HC$_3$N $J$\,=\,4--3 can be found from the broadly distributed 
    outer shells at $v_z\ga10\,\kmps$.
  \item An eccentric binary orbit is hinted in the $^{12}$CO $J$\,=\,2--1 
    map by a double spiral feature at the systemic velocity and by the 
    lack of interarm emission to the west, indicating the location of 
    orbit periastron. 
    The pattern spacing changes from $\sim2\farcs5$ between the outermost 
    arms beyond $\sim7\arcsec$ to $\sim1\farcs5$ between the inner arms, 
    characterizing an either reduced orbital period or slower wind flow. 
\end{itemize}

\section{CONCLUSION}\label{sec:ccs}

Our $^{12}$CO $J$\,=\,2--1 maps reveal a circumstellar spiral-shell pattern 
and a nascent bipolar outflow, providing an observational support for the 
binary system scenario for CIT 6. In addition, the data reveal an inner 
double spiral and interarm gaps in one direction as possible evidences for 
an eccentric binary. The eccentric orbital motion of a wide binary could 
facilitate the formation of a necessary accretion disk surrounding the 
companion to trigger a bipolar outflow emerging in the AGB-pPN transition.

\acknowledgments
We thank all SMA and SMT staffs for supporting the observations, and 
the anonymous referee for helpful suggestions to improve this paper.
H.K. acknowledges support through EACOA Fellowship from the East Asia Core 
Observatories Association, which consists of the National Astronomical 
Observatories of China, the National Astronomical Observatory of Japan, 
the Korea Astronomy and Space Science Institute, and the Academia Sinica 
Institute of Astronomy and Astrophysics. 
F.K. is supported by the Ministry of Science and Technology (MoST) of Taiwan, 
under grant numbers MOST103-2112-M-001-033- and MOST104-2628-M-001-004-MY3.

\begin{deluxetable}{ccccc}
\tablecolumns{4}
\tablecaption{SMA observation summary\label{tab:sma}}

\tablehead{\colhead{Date} & \colhead{Configuration} & \colhead{Integration 
    (hour)} & \colhead{$T_{\rm sys}$ (K)}}
\startdata
2013-01-08 & extended      & 4.4 & 70--140\\
2013-02-06 & very extended & 3.6 & 70--140\\
2013-12-12 & compact       & 3.4 & 90--160\\
2014-02-09 & subcompact    & 0.8 & 80--150\\
2014-03-11 & subcompact    & 3.8 & 90--140\\
\enddata
\end{deluxetable}

\begin{deluxetable}{cccccc}
\tablecolumns{6}
\tablecaption{SMT grid-mapping observation summary\label{tab:smt}}

\tablehead{\colhead{Molecule} & \colhead{Date} & \colhead{Grid} 
  & \colhead{Spacing (\arcsec)} & \colhead{Integration (min)} 
  & \colhead{$T_{\rm sys}$ (K)}}
\startdata
$^{12}$CO(2--1) 
& 2015-01-07 & $9\times9$ & 15 & $3\times81$ & 160--190\\
\hline
$^{13}$CO(2--1) 
& 2015-01-07 & $9\times9$ & 15 & $3\times81$ & 170--210\\
& 2015-01-17 & $7\times7$ & 15 & $3\times49$ & 180--280\\
\enddata
\end{deluxetable}

\begin{deluxetable}{cccccccc}
\tablecolumns{8}
\tablecaption{Continuum $uv$-fitting results\label{tab:con}}

\tablehead{\colhead{$\nu$} & \colhead{$uv$ coverage}
  & \colhead{Model} & \colhead{$F_\nu$}
  & \colhead{Major axis} & \colhead{Minor axis}
  & \colhead{$\Delta$RA\tablenotemark{a}}
  & \colhead{$\Delta$Dec\tablenotemark{a}}\\
\colhead{(GHz)} & \colhead{(k$\lambda$)}
  & \colhead{} & \colhead{(mJy)} 
  & \colhead{(\arcsec)} & \colhead{(\arcsec)}
  & \colhead{(mas)} & \colhead{(mas)}}
\startdata
226.2 & 4\,--\,384 & Point +   & 26.4\,$\pm$\,0.5 
      &                 &                   &  0\,$\pm$\,5  &  0\,$\pm$\,4 \\
      &         & Gaussian & 34.5\,$\pm$\,1.2 
      & 3.7\,$\pm$\,0.2 & 3.0\,$\pm$\,0.2   & 20\,$\pm$\,70 & 50\,$\pm$\,63\\
\hline
\hline
350.6 & 42\,--\,596 & Point    & 77.4\,$\pm$\,2.6 
      &                 &                   & -20\,$\pm$\,6 & -10\,$\pm$\,6\\
\enddata
\tablenotetext{a}{ Measured J2000 coordinates of CIT 6 are 
10$^{\rm h}$16$^{\rm m}$02\fs259 and $+$30\arcdeg34\arcmin19\farcs18 
in RA and Dec, respectively.}
\end{deluxetable}

\begin{figure} 
  \epsscale{0.7}
  \plotone{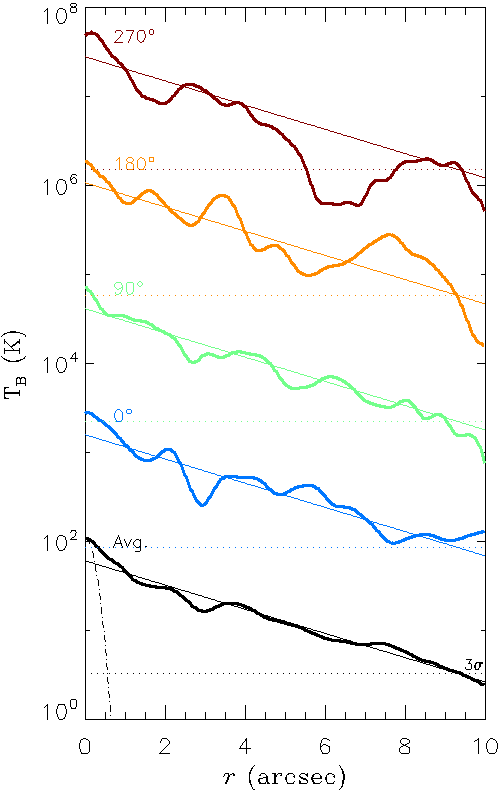}
  \caption{\label{fig:prf}
    Radial profiles of $^{12}$CO $J$\,=\,2--1 at the systemic velocity (width 
    of 5\,\kmps). Bottom black curve presents the azimuthally averaged profile 
    over 360\arcdeg, overlaid by a thin line representing a linear fit (slope 
    of $-0.31$) in the log-linear scale (i.e., an exponential function). 
    Dotted and dash-dotted lines indicate the $3\sigma$ level and the 
    synthesized Gaussian beam, respectively. 
    The upper four curves represent 
    the profiles along the noted PA averaged over 20 degree-width sectors. 
    They are vertically shifted (by 3.3 in the logarithmic scale) together 
    with the thin solid and dotted lines, which denote analogous quantities 
    (averaged fit and $3\sigma$ level, respectively) as for the black curves 
    but for the different PAs. 
  }
  \epsscale{1}
\end{figure}

\begin{figure*} 
  \includegraphics[width=0.33\textwidth]{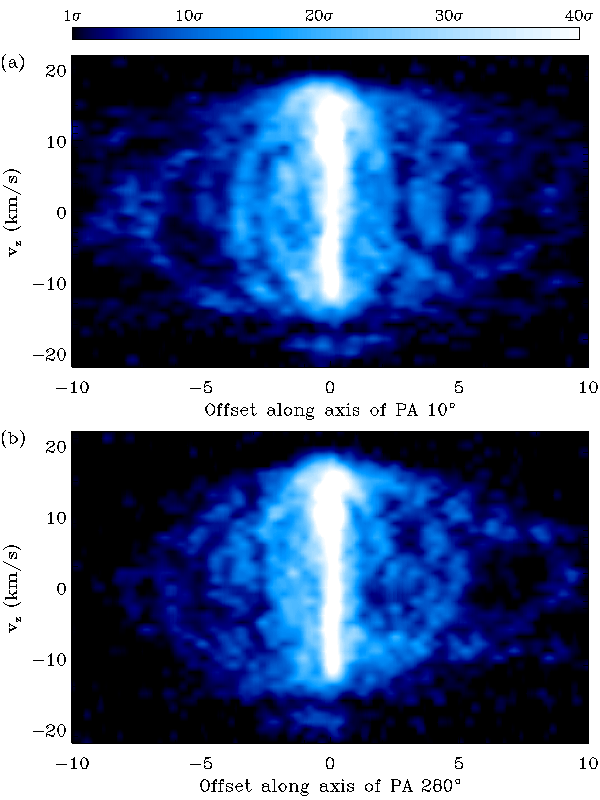}
  \includegraphics[width=0.33\textwidth]{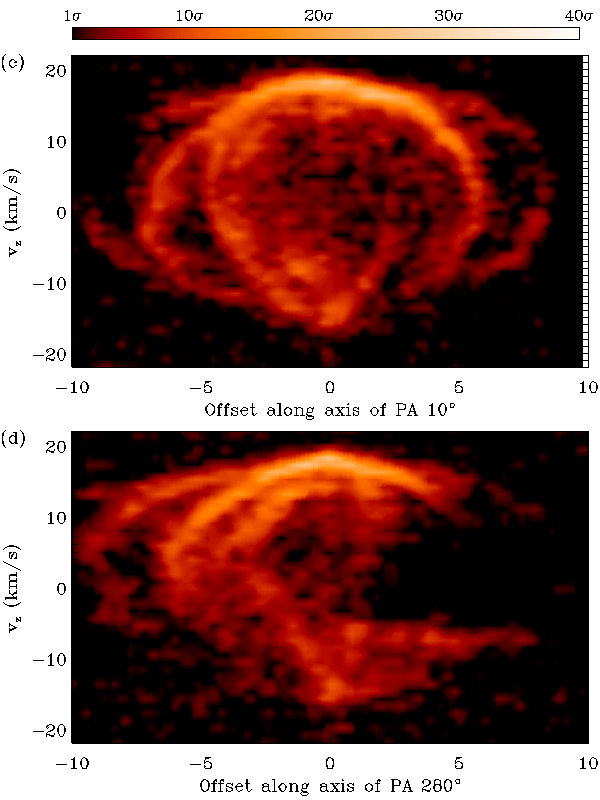}
  \includegraphics[width=0.33\textwidth]{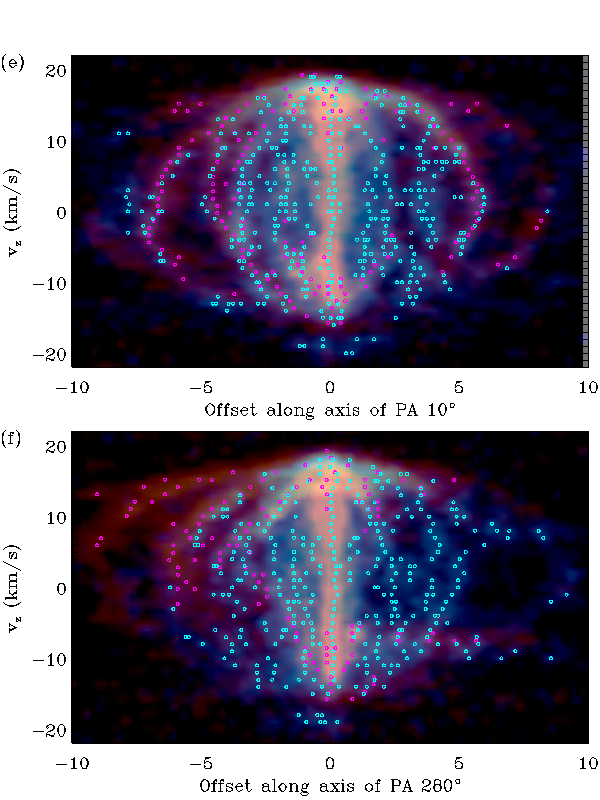}
  \caption{\label{fig:pvd}
    Position-velocity diagrams of the $^{12}$CO $J$\,=\,2--1 (blue) and 
    HC$_3$N $J$\,=\,4--3 (red) for PAs of $10\arcdeg$ (upper panels) and 
    280\arcdeg\ (lower panels). The negative offset means the opposite 
    direction, i.e., PA of $190\arcdeg$ (upper) and $100\arcdeg$ (lower). 
    The velocity interval is 1\,\kmps, and is denoted with respect to 
    the systemic velocity. The open circles in (e)--(f) refer the local 
    peak positions selected with the criteria of 10$\sigma$ for $^{12}$CO 
    ($\sigma=24$\,\mJpb) and 5$\sigma$ for HC$_3$N ($\sigma=0.5$\,\mJpb). 
  }
\end{figure*}

\begin{figure*} 
  \plotone{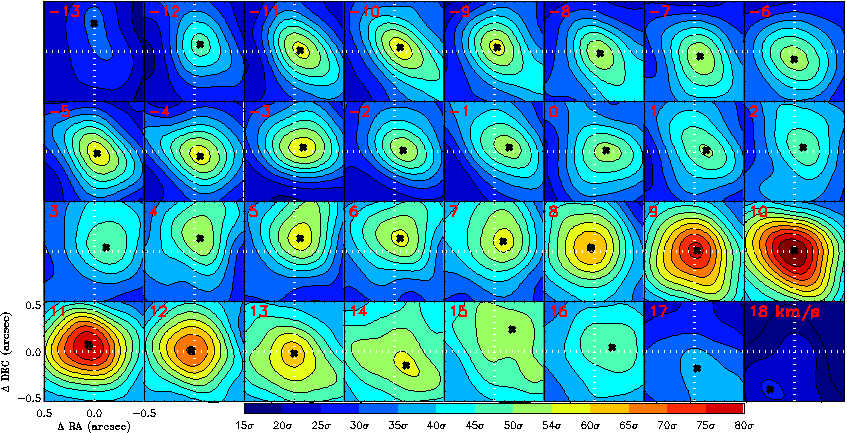}
  \caption{\label{fig:cpk}
    Channel map of $^{12}$CO $J$\,=\,2--1 in the central region of CIT 6, 
    taken with the SMA. The emission peak position of each channel is 
    marked by a cross symbol. The number at the top left corner of each 
    panel represents the center velocity of the channel with respect to 
    the systemic velocity. The colorbar and contour levels are from 
    $15\sigma$ to $80\sigma$ with a step of $5\sigma$, where the noise 
    level $\sigma=24\,\mJpb$ (2.2\,K). 
    Continuum peak position is indicated by the intersection of dotted lines.
  }
\end{figure*}

\begin{figure} 
  \plotone{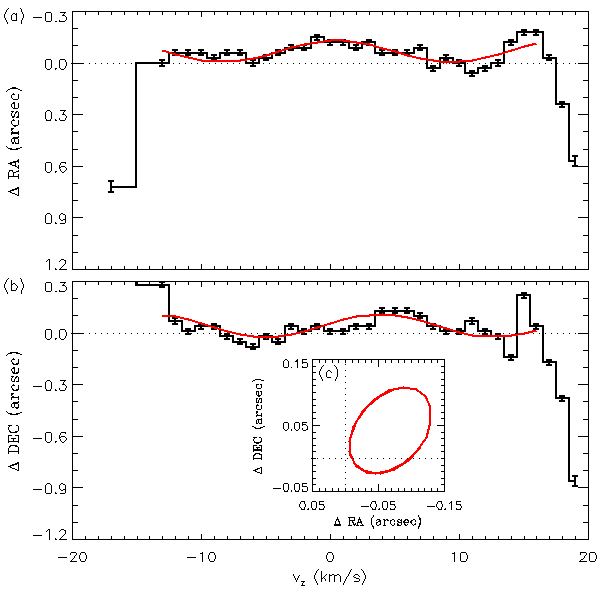}
  \caption{\label{fig:fit}
    Peak position of the $^{12}$CO $J$\,=\,2--1 line emission in each 
    channel along (a) RA and (b) Dec directions separately, with respect 
    to the coordinate center defined by the peak of the continuum emission 
    at 226.2\,GHz. The red lines show the fitting results of the offsets 
    with a functional form of $f=a_0+a_1\sin(2\pi(v_z-a_3)/a_2)$. The 
    resulting fit constants are $(a_0,a_1,a_2,a_3)=(-0.07,0.06,17.7,4.89)$ 
    for RA and $(0.04,0.07,17.7,-0.58)$ for Dec. 
    The vertical bars represent the position measurement uncertainties 
    defined by the larger values between the synthesized beam size divided 
    by the signal-to-noise ratio and the pixel size of the channel map. 
    (c) The position curve in $\Delta$RA and $\Delta$DEC are based upon 
    the fits from (a) and (b). Dotted lines are inserted to show the 
    continuum peak position. 
  }
\end{figure}

\begin{figure} 
  \plotone{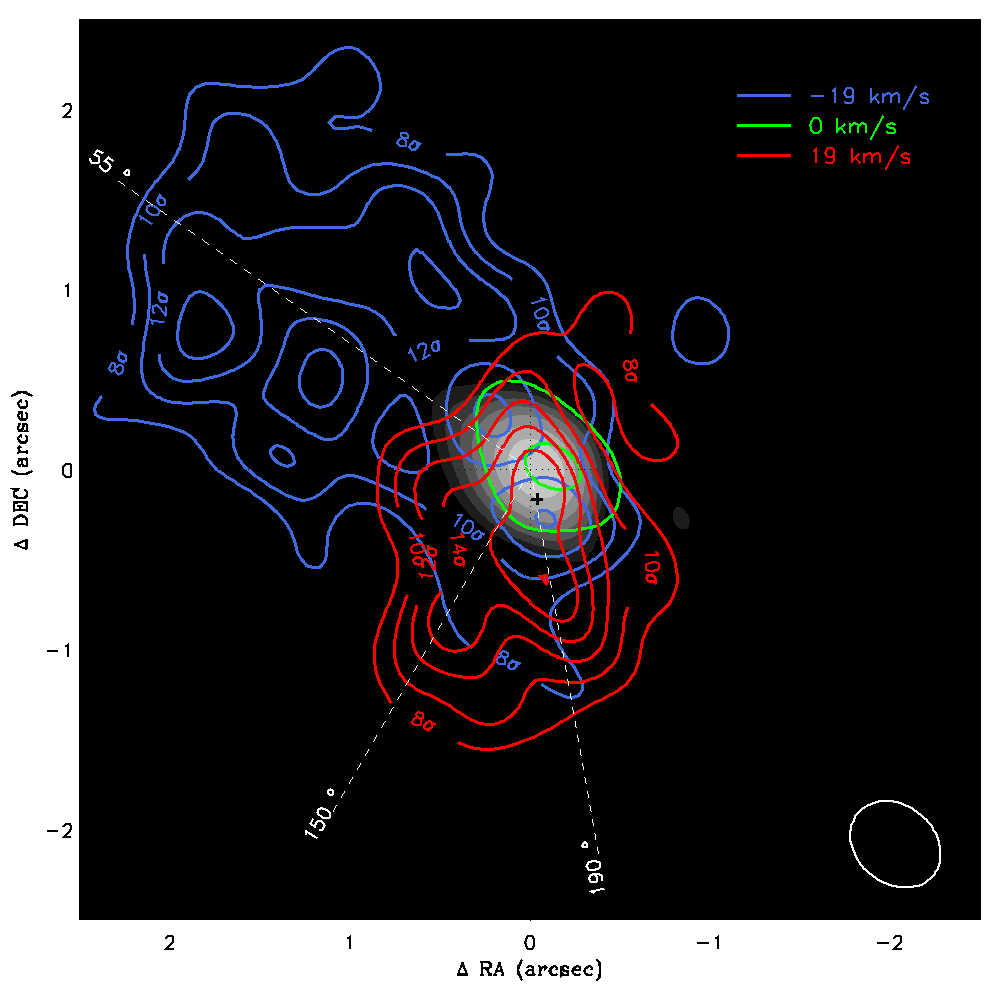}
  \caption{\label{fig:jet}
    Map of the $^{12}$CO $J$\,=\,2--1 emission averaged over $3\,\kmps$ 
    width for velocities $v_z=-19\,\kmps$ (blue contours) and $19\,\kmps$ 
    (red), compared to the contour map for the systemic velocity (green) 
    and the 226.2\,GHz continuum emission map (gray-scaled image). Three 
    position angles are marked to represent the outflow directions. The 
    cross symbol refers to the position of companion star (blue component 
    of two optical point sources in \citealp{mon00}), assuming the red 
    point source is located at the continuum peak position. 
    The contour levels are $8\sigma$, $10\sigma$, $12\sigma$, $14\sigma$, 
    $16\sigma$, and $18\sigma$ for the high-end velocity channels (blue 
    and red), and are $60\sigma$ and $80\sigma$ for the systemic velocity 
    (green). The continuum map is logarithmically scaled from $3\sigma$ 
    to $50\sigma$. The synthesized beam size is denoted by a white solid 
    ellipse at the bottom right.
  }
\end{figure}

\begin{figure} 
  \plotone{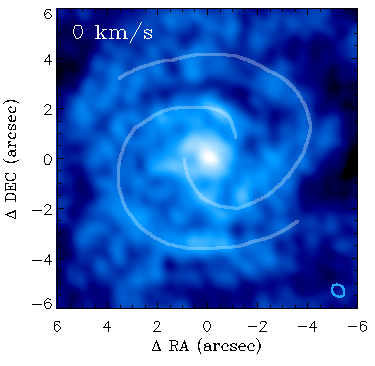}
  \caption{\label{fig:dsp}
    A double spiral feature appeared in the central region of $^{12}$CO 
    $J$\,=\,2--1 map in the systemic velocity channel. Background image 
    is the same as in the middle panel of Figure\,\ref{fig:chm}, except 
    for the smaller image size. 
  }
\end{figure}


\begin{thebibliography}{}
\bibitem[Balick \& Frank(2002)]{bal02} Balick, B., \& Frank, A.\ 2002, \araa, 40, 439 
\bibitem[Beichman et al.(1988)]{bei88} Beichman, C.~A., Neugebauer, G., Habing, H.~J., Clegg, P.~E., \& Chester, T.~J.\ 1988, Infrared astronomical satellite (IRAS) catalogs and atlases, Explanatory supplement, 1 
\bibitem[Carlstrom et al.(1990)]{car90} Carlstrom, J.~E., Welch, W.~J., Goldsmith, P.~F., \& Lis, D.~C.\ 1990, \aj, 100, 213 
\bibitem[Castro-Carrizo et al.(2010)]{cas10} Castro-Carrizo, A., Quintana-Lacaci, G., Neri, R., et al.\ 2010, \aap, 523, A59
\bibitem[Cernicharo et al.(2015)]{cer15} Cernicharo, J., Marcelino, N., Ag{\'u}ndez, M., \& Gu{\'e}lin, M.\ 2015, \aap, 575, A91 
\bibitem[Chau et al.(2012)]{cha12} Chau, W., Zhang, Y., Nakashima, J.-i., Deguchi, S., \& Kwok, S.\ 2012, \apj, 760, 66 
\bibitem[Claussen et al.(1987)]{cla87} Claussen, M.~J., Kleinmann, S.~G., Joyce, R.~R., \& Jura, M.\ 1987, \apjs, 65, 385 
\bibitem[Claussen et al.(2011)]{cla11} Claussen, M.~J., Sjouwerman, L.~O., Rupen, M.~P., et al.\ 2011, \apjl, 739, L5 
\bibitem[Cohen(1979)]{coh79} Cohen, M.\ 1979, \mnras, 186, 837 
\bibitem[Cohen \& Hitchon(1996)]{coh96} Cohen, M., \& Hitchon, K.\ 1996, \aj, 111, 962 
\bibitem[Corradi \& Schwarz(1995)]{cor95} Corradi, R.~L.~M., \& Schwarz, H.~E.\ 1995, \aap, 293, 871 
\bibitem[Decin et al.(2015)]{dec15} Decin, L., Richards, A.~M.~S., Neufeld, D., et al.\ 2015, \aap, 574, A5 
\bibitem[Dinh-V.-Trung \& Lim(2009)]{din09} Dinh-V.-Trung, \& Lim, J.\ 2009, \apj, 701, 292 
\bibitem[Groenewegen et al.(2011)]{gro11} Groenewegen, M.~A.~T., Waelkens, C., Barlow, M.~J., et al.\ 2011, \aap, 526, A162 
\bibitem[Gurwell et al.(2007)]{gur07} Gurwell, M.~A., Peck, A.~B., Hostler, S.~R., Darrah, M.~R., \& Katz, C.~A.\ 2007, in ASP Conf. Ser. 375, From Z-Machines to ALMA: (Sub)Millimeter Spectroscopy of Galaxies, ed. A.~J.~Baker, J.~Glenn, A.~I.~Harris, J.~G.~Mangum, \& M.~S.~Yun (San Francisco, CA: ASP), 234 
\bibitem[He(2007)]{he07} He, J.~H.\ 2007, \aap, 467, 1081
\bibitem[H{\"o}fner(2007)]{hof07} H{\"o}fner, S.\ 2007, Why Galaxies Care About AGB Stars: Their Importance as Actors and Probes, 378, 145 
\bibitem[Huarte-Espinosa et al.(2013)]{hua13} Huarte-Espinosa, M., Carroll-Nellenback, J., Nordhaus, J., Frank, A., \& Blackman, E.~G.\ 2013, \mnras, 1448 
\bibitem[Huggins et al.(2009)]{hug09} Huggins, P.~J., Mauron, N., \& Wirth, E.~A.\ 2009, \mnras, 396, 1805 
\bibitem[Hurley et al.(2000)]{hur00} Hurley, J.~R., Pols, O.~R., \& Tout, C.~A.\ 2000, \mnras, 315, 543 
\bibitem[Kim et al.(2013)]{kim13} Kim, H., Hsieh, I.-T., Liu, S.-Y., \& Taam, R.~E.\ 2013, \apj, 776, 86 
\bibitem[Kim \& Taam(2012a)]{kim12a} Kim, H., \& Taam, R.~E.\ 2012a, \apj, 744, 136 
\bibitem[Kim \& Taam(2012b)]{kim12b} Kim, H., \& Taam, R.~E.\ 2012b, \apj, 759, 59 
\bibitem[Kim \& Taam(2012c)]{kim12c} Kim, H., \& Taam, R.~E.\ 2012c, \apj, 759, L22 
\bibitem[Knapp et al.(1979)]{kna79} Knapp, G.~R., Kuiper, T.~B.~H., \& Zuckerman, B.\ 1979, \apj, 233, 140 
\bibitem[Koda et al.(2011)]{kod11} Koda, J., Sawada, T., Wright, M.~C.~H., et al.\ 2011, \apjs, 193, 19 
\bibitem[Lagadec et al.(2005)]{lag05} Lagadec, E., M{\'e}karnia, D., de Freitas Pacheco, J.~A., \& Dougados, C.\ 2005, \aap, 433, 553
\bibitem[Lindqvist et al.(2000)]{lin00} Lindqvist, M., Sch{\"o}ier, F.~L., Lucas, R., \& Olofsson, H.\ 2000, \aap, 361, 1036 
\bibitem[Loup et al.(1993)]{lou93} Loup, C., Forveille, T., Omont, A., \& Paul, J.~F.\ 1993, \aaps, 99, 291 
\bibitem[Maercker et al.(2012)]{mae12} Maercker, M., Mohamed, S., Vlemmings, W.~H.~T., et al.\ 2012, \nat, 490, 232 
\bibitem[Mamon et al.(1988)]{mam88} Mamon, G.~A., Glassgold, A.~E., \& Huggins, P.~J.\ 1988, \apj, 328, 797 
\bibitem[Marshall et al.(1992)]{mar92} Marshall, C.~R., Leahy, D.~A., \& Kwok, S.\ 1992, \pasp, 104, 397
\bibitem[Mastrodemos \& Morris(1999)]{mas99} Mastrodemos, N., \& Morris, M.\ 1999, \apj, 523, 357 
\bibitem[Mauron \& Huggins(1999)]{mau99} Mauron, N., \& Huggins, P.~J.\ 1999, \aap, 349, 203 
\bibitem[Mauron \& Huggins(2006)]{mau06} Mauron, N., \& Huggins, P.~J.\ 2006, \aap, 452, 257 
\bibitem[Mauron et al.(2013)]{mau13} Mauron, N., Huggins, P.~J., \& Cheung, C.-L.\ 2013, \aap, 551, A110
\bibitem[Mayer et al.(2011)]{may11} Mayer, A., Jorissen, A., Kerschbaum, F., et al.\ 2011, \aap, 531, L4 
\bibitem[Menten et al.(2012)]{men12} Menten, K.~M., Reid, M.~J., Kami{\'n}ski, T., \& Claussen, M.~J.\ 2012, \aap, 543, AA73 
\bibitem[Milam et al.(2009)]{mil09} Milam, S.~N., Woolf, N.~J., \& Ziurys, L.~M.\ 2009, \apj, 690, 837 
\bibitem[Mohamed \& Podsiadlowski(2012)]{moh12} Mohamed, S., \& Podsiadlowski, P.\ 2012, Baltic Astronomy, 21, 88 
\bibitem[Monnier et al.(2000)]{mon00} Monnier, J.~D., Tuthill, P.~G., \& Danchi, W.~C.\ 2000, \apj, 545, 957 
\bibitem[Morris(1975)]{mor75} Morris, M.\ 1975, \apj, 197, 603
\bibitem[Morris et al.(1985)]{mor85} Morris, M., Lucas, R., \& Omont, A.\ 1985, \aap, 142, 107 
\bibitem[Morris et al.(2006)]{mor06} Morris, M., Sahai, R., Matthews, K., et al.\ 2006, in IAU Symp. 234, Planetary Nebulae in our Galaxy and Beyond, ed.\ M.~J.~Barlow \& R.~H.~M{\'e}ndez (Cambridge: Cambridge Univ.\ Press), 469 
\bibitem[Neri et al.(1998)]{ner98} Neri, R., Kahane, C., Lucas, R., Bujarrabal, V., \& Loup, C.\ 1998, \aaps, 130, 1 
\bibitem[Olofsson et al.(1993)]{olo93} Olofsson, H., Eriksson, K., Gustafsson, B., \& Carlstrom, U.\ 1993, \apjs, 87, 267 
\bibitem[Ramstedt et al.(2014)]{ram14} Ramstedt, S., Mohamed, S., Vlemmings, W.~H.~T., et al.\ 2014, \aap, 570, L14 
\bibitem[Reid \& Menten(1997)]{rei97} Reid, M.~J., \& Menten, K.~M.\ 1997, \apj, 476, 327 
\bibitem[R{\"o}ser et al.(2008)]{roe08} R{\"o}ser, S., Schilbach, E., Schwan, H., et al.\ 2008, \aap, 488, 401
\bibitem[Sahai et al.(2007)]{sah07} Sahai, R., Morris, M., S{\'a}nchez Contreras, C., \& Claussen, M.\ 2007, \aj, 134, 2200 
\bibitem[Sahai et al.(1998)]{sah98} Sahai, R., Trauger, J.~T., Watson, A.~M., et al.\ 1998, \apj, 493, 301 
/\bibitem[Sahai \& Mack-Crane(2014)]{sah14} Sahai, R., \& Mack-Crane, G.~P.\ 2014, \aj, 148, 74 
\bibitem[Schmidt et al.(2002)]{sch02} Schmidt, G.~D., Hines, D.~C., \& Swift, S.\ 2002, \apj, 576, 429 
\bibitem[Sch{\"o}ier \& Olofsson(2001)]{sch01} Sch{\"o}ier, F.~L., \& Olofsson, H.\ 2001, \aap, 368, 969 
\bibitem[Sch{\"o}ier et al.(2005)]{sch05} Sch{\"o}ier, F.~L., van der Tak, F.~F.~S., van Dishoeck, E.~F., \& Black, J.~H.\ 2005, \aap, 432, 369 
\bibitem[Soker(1994)]{sok94} Soker, N.\ 1994, \mnras, 270, 774 
\bibitem[Theuns \& Jorissen(1993)]{the93} Theuns, T., \& Jorissen, A.\ 1993, \mnras, 265, 946 
\bibitem[Wood(1979)]{woo79} Wood, P.~R.\ 1979, \apj, 227, 220 
\bibitem[Zhang et al.(2009)]{zha09} Zhang, Y., Kwok, S., \& Dinh-V-Trung 2009, \apj, 691, 1660 
\end{thebibliography}
\end{document}